arXiv: Research article

# How to quantify interaction strengths? A critical rethinking of the interaction Jacobian and evaluation methods for non-parametric inference in time series analysis

**Takeshi Miki[1, 2, 3, 4], Chun-Wei Chang[3], Po-Ju Ke[4], Arndt Telschow[5], Cheng-Han Tsai[6], Masayuki Ushio[7], Chih-hao Hsieh[4,8,9,10]**

1. Faculty of Advanced Science and Technology, Ryukoku University, Japan
2. Center for Biodiversity Science, Ryukoku University, Japan
3. Institute of Fisheries Science, National Taiwan University, Taiwan
4. Institute of Ecology and Evolutionary Biology, National Taiwan University, Taiwan
5. Cuculus GmbH, Germany
6. Department of Life Sciences, National Cheng Kung University, Taiwan
7. Department of Ocean Science, Hong Kong University of Science and Technology, Hong Kong
8. Institute of Oceanography, National Taiwan University
9. National Center for Theoretical Sciences, Taiwan
10. Research Center for Environmental Changes, Academia Sinica, Taiwan

## Abstract

Quantifying interaction strengths between state variables in dynamical systems is essential for understanding ecological networks. Within the empirical dynamic modeling approach, multivariate S-map infers the interaction Jacobian from time series data without assuming specific dynamical models. This approach enables the non-parametric statistical inference of interspecific interactions through state space reconstruction. However, deviations in the biological interpretation and numerical implementation of the interaction Jacobian from its mathematical definition pose challenges. We mathematically reintroduce the interaction Jacobian using differential quotients, uncovering two problems: (1) the mismatch between the interaction Jacobian and its biological meaning complicates comparisons between interspecific and intraspecific interactions; (2) the interaction Jacobian is not fully implemented in the parametric Jacobian numerically derived from given parametric models, especially using ordinary differential equations. As a result, model-based evaluations of S-map methods become inappropriate. To address these problems, (1) we propose adjusting the diagonal elements of the interaction Jacobian by subtracting 1 to resolve the comparability problem between inter- and intraspecific interaction strengths. Simulations of population dynamics showed that this adjustment prevents overestimation of intraspecific interaction strengths. (2) We introduce an alternative parametric Jacobian and then cumulative interaction strength (CIS), providing a more rigorous benchmark for evaluating S-map methods. Furthermore, we demonstrated that the numerical gap between CIS and the existing parametric Jacobian is substantial in realistic scenarios, suggesting CIS as preferred benchmark. These solutions offer a clearer framework for developing non-parametric approaches in ecological time series analysis.

## 1. Introduction

Non-parametric approaches for understanding complex dynamical systems have been developed in the last three decades as nonlinear time series analysis for short-term forecasting (Farmer and Sidorowich, 1987; Sugihara, 1994; Sugihara and May, 1990) and for causality inference (Hirata et al., 2016; Hirata and Aihara, 2010; Munch et al., 2023; Osada et al., 2023; Schreiber, 2000; Sugihara et al., 2012). These methods aim to reconstruct an interaction network of state variables from empirical time series, without assuming any underlying dynamical equations. In contrast, parametric approaches (in sensu Song and Saavedra, 2021) assume specific dynamical models and fit them to time series data. These approaches typically use either the continuous-time framework with ordinary differential equations (ODEs) or the discrete-time framework with difference equations (DEs). When applied to population and community ecology, intra- and inter-specific interactions in non-parametric approaches are defined by the net effects on population growth. These effects include both direct and indirect influence, which are consistent with the contemporary definition of interaction in these fields (Abrams, 1987).

Empirical dynamic modeling (EDM), a suite of tools for nonlinear time series analyses developed in recent decades, enables the study of dynamical systems without assuming governing equations. Based on Takens' Theorem (Takens, 1981) and its extension (Sauer et al., 1991), EDM reconstructs the system's manifold from time series data, allowing the inference of system dynamics through state space reconstruction. Within this research program, multivariate S-map and its derived methods, infer the interactions among time series variables using a sequential locally weighted global linear map (S-map)(Deyle et al., 2016; Sugihara, 1994; see also references in **Table 1** for other variations). These algorithms have been developed into a standard non-parametric method to quantify state-dependent interaction strengths through empirically inferring state-dependent 'interaction Jacobian' matrix. As one of the core components in EDM (Munch et al., 2023), the interaction Jacobian is informative for quantifying the sign and magnitude of interspecific interactions and characterizing interaction network structures (Chang et al., 2020; Mondal et al., 2022; Natsukawa et al., 2021; Ushio et al., 2018). This is also useful for understanding the system properties such as dynamical stability, which is the ability of a system to resist divergence against perturbation on system state variable (Grziwotz et al., 2023; Li and Liu, 2023; Ushio et al., 2018).

**Table 1 Summary of non-parametric methods to infer interaction Jacobian**

| **Method name and reference** | **Regression formula for the time series *Y* to obtain state-dependent linear coefficients** | **Correspondence between state-dependent linear coefficients and their mathematical notation** | **Parametric dynamical model used for benchmarking the proposed state-dependent linear regression method** |
|---|---|---|---|
| **Multivariate S-map** a) Deyle et al (2016) | $N_i(k+1) = C_0 + \sum_j C_{ij} N_j(k)$ | $C_{ij} \to \frac{\partial x_i(k+1)}{\partial x_j(k)}$ | ODE with noises (Deyle et al 2016), DE without noises (Ushio et al 2018) |
| **Sparse S-map** b) Suzuki et al (2017) | $ln[N_i(k+1)]$-$ln[N_i(k)] = C_0 + \sum_j C_{ij} N_j(k)$ | $C_{ij} \to \frac{\partial}{\partial x_j(k)} \ln \frac{x_i(k+1)}{x_i(k)}$ | ODE with noises |
| **Regularized S-map** Cenci et al (2019) | $N_i(k+1) = C_0 + \sum_j C_{ij} N_j(k)$ | $C_{ij} \to \frac{\partial x_i(k+1)}{\partial x_j(k)}$ | SDE (= ODE + Weiner process) |
| **MDR S-map** Chang et al. (2021) | $N_i(k+1) = C_0 + \sum_j C_{ij} N_j(k)$ | $C_{ij} \to \frac{\partial x_i(k+1)}{\partial x_j(k)}$ | DE with noises |
| **LMDr** c) Kawatsu (2024) | $N_i(k+1) = C_0 + \sum_j C_{ij} N_j(k)$ | $C_{ij} \to \frac{\partial x_i(k+1)}{\partial x_j(k)}$ | ODE and DE with noises |
| **Adjusted interaction strength** d) This study | $N_i(k+1) = C_0 + \sum_j \left[C_{i,j}^{adj} + \delta_{ij}\right] N_j(k)$ | $C_{i,j}^{adj} \to \frac{\partial x_i(k+1)}{\partial x_j(k)} - \delta_{ij}$ | ODE and DE without noises |

Note: a) We denote the multivariate S-map as the *standard* S-map in Section 5 for simplicity. b) The sparse S-map first assumes a specific form of ODEs as $\frac{dx_i}{dt} = f_i(\boldsymbol{x})x_i$ and is intended to estimate the effect of $x_j$ on the per-capita growth rate of $x_i$ by $\frac{\partial}{\partial x_j}\frac{d}{dt}\ln x_i = \frac{\partial}{\partial x_j}\frac{1}{x_i}\frac{dx_i}{dt} = \frac{\partial f_i}{\partial x_j}$. c) Strictly speaking, LMDr is not merely a variant of the multivariate S-map method but introduces a new state-dependent regression method based on the local manifold distance (LMD). d) The adjusted interaction strength can rely on any of the existing state-dependent linear regression methods, except for the sparse S-map. Also note that the state-dependent linear regression refers to S-map or LMDr and the state-dependent linear coefficients refer to S-map coefficients or coefficients of LMDr model.

The interaction Jacobian (in sensu Chang et al. 2021) is defined as the Jacobian of the flow (Deyle et al., 2016; Munch et al., 2022). However, some inconsistency and ambiguity regarding interaction Jacobian hinder further development and application of S-map methods for ecological dynamics. There is a gap between the mathematical definition of the interaction Jacobian and its biological interpretation. On the one hand, the interaction Jacobian in a dynamical system with $n$ dimensional state vector $x$ is explicitly and uniquely defined as the matrix of partial derivatives of the $n$-dimensional vector-valued function that maps from $x(t)$ to $x(t+\tau)$, where $x(t)$ and $x(t+\tau)$ represent the system state at time $t$ and $t+\tau$, respectively (Deyle et al., 2016; Munch et al., 2022). In other words, the interaction Jacobian describes how the pulse disturbance at time $t$ affects the system state at time $t+\tau$. Then, the interaction strengths in the non-parametric approaches are operationally defined as the elements of the interaction Jacobian. On the other hand, the intuitive interpretation of interaction Jacobian for linking it to interaction strengths has been described in ambiguous and inconsistent ways such as, *net local effect* of each of state variables on the target variable (Deyle et al., 2016), *the population-level interaction between two species* (Ushio et al., 2018), *the change in growth of a species as a result of a change in abundance of another species* (Cenci et al., 2019), or *the net effects of abundance changes in nodes j (between two consecutive observations) on the abundance of species i* (Chang et al., 2021).

The ambiguity and inconsistency stem from the fact that most previous research has been published in journals intended primarily for biologists, where highly mathematical statements are often avoided. As a result, two key problems arise: one biological, the other mathematical. As the first, biological problem, it is unclear whether the interaction Jacobian fits well with the biological concept of interaction strengths because of the operationality in defining interaction strengths despite varying interpretations (Cenci et al., 2019; Chang et al., 2021; Deyle et al., 2016; Ushio et al., 2018). We will later demonstrate that the diagonal element of the interaction Jacobian does not fit well with the biological concept of intraspecific interaction strength.

The second, mathematical problem is whether the uniquely defined interaction Jacobian is mathematically equivalent to the Jacobian coefficients derived from the mathematical formulation in parametric population dynamics models. This equivalence is critical, as the latter serves as a benchmark for method development. Hereafter, we will use the term *parametric* Jacobian when referring to the Jacobian matrix numerically calculated from parametric models, addressing the role of parametric models in developing and validating new methods. The conventional strategy to evaluate S-map methods is to first generate time series data from a parametric population dynamics model and apply the S-map method to infer the interaction Jacobian (Chang et al., 2021; Deyle et al., 2016). The second step is to calculate the parametric Jacobian from the original data-generating model as a benchmark and compare it with the inferred interaction Jacobian. When using data simulated from a discrete-time model framework (DEs), it is easily confirmed that the parametric Jacobian evaluated at a focal time step is mathematically equivalent to the interaction Jacobian, because it fully captures how the system changes at the next time step. However, in a continuous-time model framework (ODEs), the parametric Jacobian evaluated at a focal time $t$ does not account for the continuous changes in state variables between $t$ and $t+\tau$, and thus cannot capture their cumulative effects on the state variables at $t+\tau$. We will later demonstrate that this results in a discrepancy between the parametric Jacobian and the interaction Jacobian, implying that the simple parametric Jacobian might not effectively serve as a benchmark for the interaction Jacobian.

To resolve the ambiguity and inconsistency, we mathematically reintroduce the interaction Jacobian as the matrix of partial derivatives of the $n$-dimensional vector-valued function that maps from $x(t)$ to $x(t+\tau)$, without relying on the conventional partial derivative notation $\frac{\partial x_i(t+\tau)}{\partial x_j(t)}$. Instead, we start our derivation with differential quotients to provide greater transparency in both the mathematical formulation and its biological interpretation. Disclosing implicit aspects in EDM (Edwards et al., 2024) will enhance our understanding of the interaction Jacobian. With this exercise, we are able to determine whether the interaction Jacobian are consistent with biological intuition on interaction strengths and correct any aspects that are not. This will be achieved through deriving the interaction Jacobian when there are no net dynamical changes in the system from time $t$ to $t+\tau$. With this reintroduction, we aim to explicitly reframe the S-map-based methods for quantifying interaction strengths and then to propose an adjusted metric of interaction strengths based on the interaction Jacobian and S-map coefficients.

The reintroduced definition of the interaction Jacobian also enables us to derive the relationship between the interaction Jacobian and the parametric Jacobian from an ODE model. With this derivation, we aim to highlight the limitations of using the parametric Jacobian as a benchmark for evaluating the performance of S-map-based methods in any dataset obtained from ODE systems. More specifically, we will demonstrate that the commonly used parametric Jacobian, evaluated at a single time point, is only an approximation, as it neglects the continuity of system state changes, and will also show when the approximation becomes ineffective. Finally, we propose mathematical and numerical methods to calculate an alternative parametric Jacobian that includes cumulative effects of continuous changes of system variables. This would serve as a more suitable benchmark when assessing S-map algorithms by simulated data from an ODE system.

We introduce the mathematical framework in Section 2. In Section 3, we reintroduce the interaction Jacobian starting from previously skipped mathematical formulation, using differential quotients, and identify the two problems argued above. In Section 4 we propose the solutions for these problems while presenting two applications in Section 5. We do not intend to evaluate the relative performance of different S-map algorithms by using the proposed new methodology. Rather, with a couple of examples, we demonstrate the further space for improving the S-map algorithms in Section 5. In Section 6, we discuss our findings and their implications for ecological application, method development, and mathematicians, including cautions about when to use the original interaction Jacobian and when to use the adjusted metric for interaction strengths. Mathematical notations, symbols, and quantities used in the following sections are summarized in **Table 2**.

**Table 2: List of notations, symbols, and quantities**

| |
|---|
| DE: Difference equations for a discrete-time population dynamics model |
| ODE: Ordinary differential equations for a continuous-time population dynamics model |
| $\mathbf{Y}$: The discrete-time observations |
| $t$: Time in a continuous-time framework. |
| $\tau$: The interval (> 0) between two datapoints of $\mathbf{Y}$, which also represents the actual time interval between two steps for the DE framework. |
| $k$: The $k$-th data point ($k$ = 1, 2, …) when used for discrete-time observations $\mathbf{Y}$, which is also used for the $k$-th time step of the DE framework ($k$ = 0, 1., 2,…), and for ODE framework as $t = k\tau$ ($k$ = 0, 1, 2,…). |
| $\boldsymbol{N}(k)$: The $n$-dimensional observation vector at the $k$-th data point, $k$ = 1, 2, …. |

$\boldsymbol{x}(k)$: The $n$-dimensional state vector for the DE framework at $k$-th time temp, $k = 0, 1, 2, \ldots$

$\boldsymbol{x}(t)$: The $n$-dimensional state vector for the ODE framework at time $t \geq 0$.

$\boldsymbol{F}$: The $n$-dimensional vector valued function that characterizes a dynamical equation in the DE framework.

$\boldsymbol{f}$: The $n$-dimensional vector valued function that characterizes a dynamical equation in the ODE framework.

$\phi(\boldsymbol{x}, k)$: The flow (mapping $\phi: \mathbb{R}^n \times \mathbb{Z} \to \mathbb{R}^n$) in a discrete dynamical system, which satisfies $\phi(\boldsymbol{x}_0, 0) = \boldsymbol{x}_0$ and $\phi(\phi(\boldsymbol{x}_0, k), l) = \phi(\boldsymbol{x}_0, k + l)$.

$\phi_i(\boldsymbol{x}, t)$: The $i$-th element of the flow, $i = 1, 2, \ldots, n$.

$\varphi(\boldsymbol{x}, t)$: The flow (mapping $\varphi: \mathbb{R}^n \times \mathbb{R} \to \mathbb{R}^n$) in a continuous dynamical system, which satisfies $\varphi(\boldsymbol{x}_0, 0) = \boldsymbol{x}_0$ and $\varphi(\varphi(\boldsymbol{x}_0, t), s) = \varphi(\boldsymbol{x}_0, t + s)$.
$\varphi_i(\boldsymbol{x}, t)$: The $i$-th element of the flow, $i = 1, 2, \ldots, n$.

$\Delta\boldsymbol{x}_j$: The $n$-dimensional small perturbation on the $j$-th element, defined more specifically as $\Delta\boldsymbol{x}_j = \left(\underbrace{0,0,\ldots 0}_{j-1}, \Delta x, \underbrace{0\ldots,0}_{n-j}\right)$.

$\delta_{ij}$: Kronecker delta, defined as $1$ if $i = j$, and $0$ if $i \neq j$. Using this, the $i$-th element of $\Delta\boldsymbol{x}_j$, is given by $[\Delta\boldsymbol{x}_j]_i = \delta_{ij}\Delta x$.

$\mathbf{I}_n \in \mathbb{R}^{n \times n}$: The identity matrix, where the $(i, j)$-th element is $\delta_{ij}$.

$\mathbf{J}$: The interaction Jacobian, mathematically defined as the Jacobian of the flow in both the DE and ODE frameworks. This is the quantity that the multivariate S-map methods and other non-parametric methods aim to infer from an observed time series as well as a simulated time series.

$\mathbf{S}^{adj}$: The matrix of the adjusted interaction strength, mathematically defined in both the DE and ODE frameworks, with the relationship that $\mathbf{J} = \mathbf{S}^{adj} + \mathbf{I}_n$.

Parametric Jacobian: The interaction Jacobian numerically derived from a specific parametric population dynamics model, which serves as either an approximation of or an exact equivalent to the interaction Jacobian, depending on the definition.

Parametric interaction strength: The adjusted interaction strength, defined as the parametric Jacobian minus $\delta_{ij}$.

$\mathbf{S}^{I,adj}$: Instantaneous interaction strength (IIS), referring to the parametric interaction strength for the ODE framework, derived from the parametric Jacobian serving as an approximation of the interaction Jacobian, with the relationship that $\mathbf{J} \approx \mathbf{S}^{I,adj} + \mathbf{I}_n$.

$\mathbf{S}^{C,adj}$: Cumulative interaction strength (CIS), referring to the parametric interaction strength for the ODE framework, derived from the parametric Jacobian exactly equivalent to the interaction Jacobian, with the relationship that $\mathbf{J} = \mathbf{S}^{C,adj} + \mathbf{I}_n$.

## 2. Mathematical framework for time series analysis and parametric models

### *2.1. Definition of the observations and basic assumptions*

Consider a situation where we have discrete-time observations, i.e., the time series data, for a *n*-dimensional system with the interval $\tau$:

$$\mathbf{Y} = (\boldsymbol{N}(1), \dots, \boldsymbol{N}(k), \dots, ), \boldsymbol{N}(k) \in \mathbb{R}^n, \tag{1}$$

where *k* denotes the *k*-th data point at time $t = k\tau$ in the continuous-time framework. In general, the *n* state variables can include both biological and abiotic components, but here we simply refer the time series data to multi-species population dynamics data. Here the index *k* does not imply the *k*-th generations in the context of population dynamics. Similarly, the time interval $\tau$ can be operationally chosen and is not necessarily comparable to the generation time of a population.

Here, it is reasonable to assume that we do not know how the time series $\mathbf{Y}$ was generated but that it was generated at least partly by deterministic driving forces. In other words, we can reasonably assume that $\mathbf{Y}$ was generated by either difference equations (DEs) or ordinary differential equations (ODEs) with or without noises. However, it is fundamentally impossible to determine whether DEs or ODEs were the driving forces, and their exact mathematical formulations remain unknown. In this situation, to infer the underlying dynamics of the time series data, we must rely on methods capable of providing algorithms and metrics that are compatible with both DE and ODE frameworks. In this study, we refer to such methods as non-parametric methods.

### *2.2. Parametric population dynamics model with discrete-time framework*

In the case of a DE system, we derive an *n*-dimensional DEs starting from the *1st principle formulation* (Gotelli, 2001) considering birth and death processes (but neglecting spatial processes for simplicity):

$$\boldsymbol{x}(k+1)-\boldsymbol{x}(k)=\underbrace{\boldsymbol{B}\big(\boldsymbol{x}(k)\big)}_{birth}-\underbrace{\boldsymbol{D}\big(\boldsymbol{x}(k)\big)}_{death},\boldsymbol{x}(0)=\boldsymbol{x}_0.$$

This can be simplified as:

$$\boldsymbol{x}(k+1)=\boldsymbol{x}(k)+\boldsymbol{F}\big(\boldsymbol{x}(k)\big),\boldsymbol{x}(0)=\boldsymbol{x}_0. \tag{2}$$

More specifically, it can be written as:

$$x_i(k+1)=x_i(k)+F_i\big(x_1(k),x_2(k),\dots,x_n(k)\big),x_i(0)=x_{i,0},\ i=1,2,\dots,n, \tag{2$'$}$$

where $F_i$ represents the $i$-th element of the vector-valued function $\boldsymbol{F}$.

Note that $\boldsymbol{x}(0)$ and $\boldsymbol{x}_0$ represent the initial condition in the framework of the DE parametric model but do not imply the first data point of the empirical time series, which would be $\boldsymbol{N}(1)$ following our notation in Eq. (1).

This formulation differs from a phenomenological modeling approach using a recursion map, which is more commonly used in the fields of entomology and fisheries management to describe a recruitment curve ($\boldsymbol{x}(k)$ vs $\boldsymbol{x}(k+1)$). However, we employ the formulation in Eq. (2) to ensure the consistency with the formulation from an ODE system (see Eq.(4)). Consequently, the function $\boldsymbol{F}$ represents the population growth rate. Also note that the actual time interval between the sequential time steps ($k$, and $k+1$) is $\tau$. Although it may be common in theoretical studies to set the time interval equal to the generation time, it is more appropriate to assume no direct linkage between the two for better correspondence between the empirical time series data given by Eq.(1) and a parametric population dynamics model. It follows that the function $\boldsymbol{F}$ reflects both direct and indirect interactions between variables occurring in the period $\tau$. Based on this notation, we do not address the differences between direct and indirect interactions in the following sections since the indirect interactions' impact is inevitably included in both DE and ODE systems.

The system state at time step $k$, $\boldsymbol{x}(k)$, starting from the initial condition $\boldsymbol{x}_0$ at $k = 0$ and being governed by Eq. (2) or (2′) is expressed as $\phi(\boldsymbol{x}_0,k)$, using the notation of flow ($\phi$: $\mathbb{R}^n\times\mathbb{Z}\to\mathbb{R}^n$). From Eqs. (2) and (2′), we trivially obtain the following relationships, respectively:

$$\phi(\boldsymbol{x}_0,k+1)=\phi(\boldsymbol{x}_0,k)+\boldsymbol{F}\big(\phi(\boldsymbol{x}_0,k)\big), \tag{3}$$

and

$$\phi_i(\boldsymbol{x}_0,k+1)=\phi_i(\boldsymbol{x}_0,k)+F_i\big(\phi(\boldsymbol{x}_0,k)\big),\ i=1,2,\dots,n. \tag{3$'$}$$

This can be rewritten as follows:

$$\phi_i(\phi(\boldsymbol{x}_0,k),1)=\phi_i(\boldsymbol{x}_0,k)+F_i\big(\phi(\boldsymbol{x}_0,k)\big),\ i=1,2,\dots,n, \tag{3''}$$

where $\phi_i$ represents the *i*-th element of the system state (= $x_i$).

*2.3. Parametric population dynamics model with continuous-time framework*

In the case of an ODE system, we derive an *n*-dimensional ODEs starting from the 1st principle formulation (Gotelli, 2001) considering birth and death processes (but neglecting spatial processes for simplicity):

$$\frac{d\boldsymbol{x}(t)}{dt}=\underbrace{\boldsymbol{B}\big(\boldsymbol{x}(t)\big)}_{birth}-\underbrace{\boldsymbol{D}\big(\boldsymbol{x}(t)\big)}_{death},\boldsymbol{x}(0)=\boldsymbol{x}_0,\boldsymbol{x}\in\mathbb{R}^n,$$

where $\boldsymbol{x}(t)$, $\boldsymbol{B}$ and $\boldsymbol{D}$ represent the *n*-dimensional system state, the function specifying the state-dependent birth rate and that for death rate, respectively. This can be simplified using a general notation $\boldsymbol{f}$ for the demographic processes as:

$$\frac{d\boldsymbol{x}(t)}{dt}=\boldsymbol{f}\big(\boldsymbol{x}(t)\big),\boldsymbol{x}(0)=\boldsymbol{x}_0. \tag{4}$$

When explicitly writing for each element of the *n*-dimensional vector $\boldsymbol{x}(t)$, we have:

$$\frac{dx_i(t)}{dt}=f_i\big(x_1(t),x_2(t),\dots,x_n(t)\big),x_i(0)=x_{i,0},\ i=1,2,\dots,n. \tag{4'}$$

where $f_i$ represents the *i*-th element of the vector-valued function $\boldsymbol{f}$.

Here, as the same as the DE model, the system state at time *t*, $\boldsymbol{x}(t)$, starting from the initial condition $\boldsymbol{x}_0$ at $t = 0$ and being governed by Eq. (4) or (4′) is also expressed as $\varphi(\boldsymbol{x}_0,t)$, using the notation of flow ($\varphi$: $\mathbb{R}^n\times\mathbb{R}\to\mathbb{R}^n$). Then, we have the following relationship from Eq. (4):

$$\varphi(\boldsymbol{x}_0,(k+1)\tau)=\varphi(\boldsymbol{x}_0,k\tau)+\int_{k\tau}^{(k+1)\tau}\boldsymbol{f}\big(\varphi(\boldsymbol{x}_0,s)\big)ds \tag{5}$$

More specifically, it can be written as:

$$\varphi_i(\boldsymbol{x}_0,(k+1)\tau)=\varphi_i(\boldsymbol{x}_0,k\tau)+\int_{k\tau}^{(k+1)\tau}f_i(\varphi(\boldsymbol{x}_0,s))ds,\ i=1,2,\dots,n, \tag{5'}$$

where $\varphi_i$ represents the *i*-th element of the system state (= $x_i$).

Here, to have better compatibility with the DE framework given in Eq.($3''$), Eq.($5'$) can be further rewritten as:

$$\varphi_i(\varphi(\boldsymbol{x}_0, k\tau), \tau) = \varphi_i(\boldsymbol{x}_0, k\tau) + \int_0^{\tau} f_i(\,\varphi(\varphi(\boldsymbol{x}_0, k\tau), s))ds,\ i = 1,2, \dots, n. \tag{5''}$$

## 3. Reintroducing interaction Jacobian and identifying problems

### *3.1. Mathematical reintroduction of interaction Jacobian*

To ensure clarity in our study, we start this section by defining the interaction Jacobian $\mathbf{J} \in \mathbb{R}^{n\times n}$ for a dynamical system with *n* dimensional state vector $x$. **J** is uniquely defined as the matrix of partial derivatives of the *n*-dimensional vector-valued function that maps from $x(t)$ to $x(t+\tau)$, where $x(t)$ and $x(t+\tau)$ represent the system state at time *t* and $t+\tau$, respectively, with $\tau > 0$ (Deyle et al., 2016; Munch et al., 2022). In other words, **J** is the Jacobian of the flow of a dynamical system. This is the common and unique definition for both the continuous-time framework (in other words, continuous dynamical systems) and the discrete-time framework (discrete dynamical systems). Although the notation of time in this definition follows the continuous-time presentation, it can be easily translated into the discrete-time presentation. The symbol $x(t)$ is used only for this definition and is translated into $\boldsymbol{x}(k)$ and $\boldsymbol{x}(t)$ for DE and ODE framework, respectively, in the following sections. Note that the interaction Jacobian is not equivalent to the Jacobian of the right-hand side (r.h.s.) of the parametric population dynamics models (the function $\boldsymbol{F}$ in Eq. (3) and the function $\boldsymbol{f}$ in Eq. (4)).

Following the definition above, the (*i, j*)-th element of the interaction Jacobian $\mathbf{J}(k)$ in the DE framework at the time step *k* is reintroduced with the differential quotients as:

$$J_{i,j}(k) = \lim_{\Delta x \to 0} \frac{\phi_i(\boldsymbol{x}(k) + \Delta\boldsymbol{x}_j, 1) - \phi_i(\boldsymbol{x}(k), 1)}{\Delta x}, \tag{6a}$$

where $\Delta\boldsymbol{x}_j = \left(\underbrace{0,0, \dots 0}_{j-1}, \Delta x, \underbrace{0\dots,0}_{n-j}\right)$. Note that the formulation in Eq. (6a) exactly follows the definition expressed in Introduction and at the beginning of this subsection, since $\phi(\bullet\,,1)$ represents the map from $\boldsymbol{x}(k)$ to $\boldsymbol{x}(k+1)$. In **Table 1**, except for (Suzuki et al., 2017), all the authors interpreted the interaction Jacobian as the interaction strength $S_{i,j}(k)$, i.e., the impact of species *j* on species *i,* at the time step *k*, and it is explicitly written as:

$$S_{i,j}(k) \equiv J_{i,j}(k). \tag{6b}$$

In simpler terms, Eqs. (6a) and (6b) state that the interaction strength is the ratio of two quantities: 1) the change of the recipient variable $x_i$ at the time step $k + 1$ in response to a very small perturbation on the donor variable $x_j$ ($\Delta \boldsymbol{x}_j$) at the time step $k$, and 2) the size of perturbation added ($\Delta x$). More formally, and without ambiguity, the interaction strength can be described as the *marginal* change in the recipient variable $x_i$ at the time step $k + 1$ in responses to a unit-sized perturbation on the donor variable $x_j$ at the time step $k$ (Munch et al., 2023). Eqs. (6a) and (6b) are also conventionally expressed by a simple partial derivative notation (Deyle et al., 2016; Ushio et al., 2018):

$$S_{i,j}(k) = J_{i,j}(k) = \frac{\partial x_i(k+1)}{\partial x_j(k)}. \tag{6c}$$

We can also reintroduce the interaction Jacobian and the interaction strength with the ODE framework by modifying Eq. (6a) as:

$$\begin{aligned} S_{i,j}(k\tau) \equiv J_{i,j}(k\tau) &= \lim_{\Delta x \to 0} \frac{\varphi_i(\boldsymbol{x}(k\tau) + \Delta \boldsymbol{x}_j, \tau) - \varphi_i(\boldsymbol{x}(k\tau), \tau)}{\Delta x}, \\ &= \lim_{\Delta x \to 0} \frac{\varphi_i(\varphi(\boldsymbol{x}_0, k\tau) + \Delta \boldsymbol{x}_j, \tau) - \varphi_i(\varphi(\boldsymbol{x}_0, k\tau), \tau)}{\Delta x}, \end{aligned} \tag{7}$$

where $\varphi(\bullet, \tau)$ represents the map from $\boldsymbol{x}(t)$ to $\boldsymbol{x}(t + \tau)$.

### *3.2. Identification of the gap between interaction Jacobian and interaction strengths*

Consider the situation when there are no *net* dynamical changes occurring in the recipient variable $x_i$ between the time step $k$ and $k + 1$ in the DE framework given by Eq. ($3''$) or between $t = k\tau$ and $t = (k+1)\tau$ in the ODE framework given by Eq. ($5''$) after *any* small perturbations $\Delta \boldsymbol{x}_j, j = 1,2, \dots, n$ (including $\Delta \boldsymbol{x}_j = \boldsymbol{0}$).

This is realized when the following conditions are satisfied for the DE and ODE framework, respectively:

$$F_i(\phi(\boldsymbol{x}_0, k) + \Delta \boldsymbol{x}_j) = 0, \forall j = 1,2, \dots, n, \tag{8}$$

and

$$\int_0^{\tau} f_i(\,\varphi(\varphi(\boldsymbol{x}_0, k\tau) + \Delta\boldsymbol{x}_j, s))ds = 0, \forall j = 1,2, \dots, n. \tag{9}$$

When the conditions in Eqs. (8) and (9) are satisfied, we have the following interaction Jacobian by simple calculations (**Appendix 1.1**):

$$J_{i,j}(k) = \delta_{ij}\ , \tag{10}$$

and

$$J_{i,j}(k\tau) = \delta_{ij}\ , \tag{11}$$

for DE and ODE systems, respectively, where $\delta_{ij}$ is the Kronecker delta.

The identity matrix ($\mathbf{I}_n$) as the form of the interaction Jacobian, independently of the modeling frameworks (Eqs. (10) and (11)), is mathematically reasonable since the propagation of any perturbation ($\Delta\boldsymbol{x}$) under no net dynamical changes can be expressed as $\mathbf{I}_n\Delta\boldsymbol{x} = \Delta\boldsymbol{x}$ for both of the DE and ODE systems.

**Problem 1:** However, the conclusions in Eqs. (10) & (11) are biologically counterintuitive since biologists would naturally expect that all of the interaction strengths (the impact of species *j* ($\neq i$) on species *i*, and the impact of species *i* on species *i* itself) are zero when there are no net dynamics occurring (Eqs. (8) and (9)). We denote this gap between mathematical validity and biological intuition as Problem 1 in this study.

### *3.3. Identification of the gap between interaction Jacobian and parametric Jacobian*

Some studies used datasets generated from ODE-based models (Cenci et al., 2019; Deyle et al., 2016) and used the parametric Jacobian elements at the focal time *t* as a benchmark for evaluating the proposed S-map coefficient methods. Let us confirm if this is mathematically reasonable or not. Here we start with Eqs. ($5''$) and (7) and consider the approximation that that ***f*** does not change with time from $t = k\tau$ to $(k+1)\tau$. That is:

$$f_i\big(\varphi(\bullet, t)\big) \approx f_i\big(\varphi(\bullet\ ,0)\big), \qquad k\tau \le t \le (k+1)\tau. \tag{12}$$

With simple calculation under the assumption in Eq. (12) (**Appendix 1.2**), we have the interaction Jacobian given by:

$$J_{i,j}(k\tau) \approx \delta_{ij} + \tau \frac{\partial f_i}{\partial x_j}\bigg|_{t=k\tau}. \tag{13}$$

**Problem 2:** Eq. (13) tells that the commonly used parametric Jacobian for benchmarking, which uses *instantaneous* Jacobian coefficients of $\boldsymbol{f}$ evaluated at a single time point $t = k\tau$ ($\frac{\partial f_i}{\partial x_j}\big|_{t=k\tau}$), is just an approximation for the interaction Jacobian $J_{i,j}$, and would be valid only when $\tau$ is small so that the assumption in Eq. (12) is acceptable. One more remark from Eq. (13) is that the interaction Jacobian is a function of the time interval $\tau$ even when using the approximation of instantaneous Jacobian coefficients. We denote this looseness as Problem 2 in this study. It arises from regarding the instantaneous Jacobian coefficients at a single time point as mathematically equivalent to the interaction Jacobian and then using them as the benchmark.

## 4. Solutions for the gaps between interaction Jacobian, interaction strengths, and parametric Jacobian

### *4.1. Solution 1: Adjusted metric of interaction strengths from interaction Jacobian*

Define the *adjusted* interaction strength as the *subsequent* marginal change in the recipient variable $x_i$ at the time step $k + 1$ in responses to a unit-sized perturbation on the donor variable $x_j$ at the time step $k$. Then, the mathematical definition of the adjusted interaction strength matrix $\mathbf{S}^{adj}(k)$ for the DE framework can be given by a simple modification of Eq. (6b):

$$S_{i,j}^{adj}(k) \equiv J_{i,j}(k) - \delta_{ij} = \lim_{\Delta x \to 0} \frac{\phi_i(\boldsymbol{x}(k) + \Delta \boldsymbol{x}_j, 1) - \phi_i(\boldsymbol{x}(k), 1)}{\Delta x} - \delta_{ij}, \tag{14}$$

noting that this adjusted definition meets our intuitive definition of zero interaction strength as $S_{i,j}^{adj}(k) = 0$, when the condition 2 (Eqs. (8) and (A1)) are satisfied.

Similarly, from Eq. (7), the adjusted interaction strength matrix $\mathbf{S}^{adj}(k\tau)$ for the ODE framework can be defined as:

$$S_{i,j}^{adj}(k\tau) \equiv J_{i,j}(k\tau) - \delta_{ij} = \lim_{\Delta x \to 0} \frac{\varphi_i(\varphi(\boldsymbol{x}_0, k\tau) + \Delta \boldsymbol{x}_j, \tau) - \varphi_i(\varphi(\boldsymbol{x}_0, k\tau), \tau)}{\Delta x} - \delta_{ij}. \tag{14'}$$

Since the definitions in Eqs. (14) and (14′) are essentially identical, it does not matter whether the observation **Y** was governed by a DE or an ODE system. This ensures that the newly-defined $\mathbf{S}^{adj}$ retains the non-parametric feature of S-map methods. Then, the adjusted interaction strength matrix $\mathbf{C}^{adj}(k)$ inferred from the local linear regression by S-map on the observation **Y** is given by (**Table 1**):

$$N_i(k+1) = C_0 + \sum_j \left[C_{i,j}^{adj}(k) + \delta_{ij}\right] N_j(k)\,. \tag{15}$$

In other words, $\mathbf{C}^{adj}(k)$ corresponds to the adjusted interaction strength matrix, either $\mathbf{S}^{adj}(k)$ or $\mathbf{S}^{adj}(k\tau)$, while the original S-map coefficient matrix $\mathbf{C}(k) = \mathbf{C}^{adj}(k) + \mathbf{I}_n$ corresponds to the interaction Jacobian $\mathbf{J}(k)$ or $\mathbf{J}(k\tau)$ (see also **Figure. 5** in Discussion).

*4.2. Solution 2: Parametric Jacobians and cumulative interaction strengths*

For the DE system, which requires simpler considerations than the ODE system, we use Eqs. (3′) and (14) along with Taylar expansion to obtain the parametric interaction strength as:

$$S_{i,j}^{adj}(k) =$$

$$\lim_{\Delta x \to 0} \frac{\phi_i(\boldsymbol{x}_0,k) + [\Delta \boldsymbol{x}_j]_i + F_i(\phi(\boldsymbol{x}_0,k) + \Delta \boldsymbol{x}_j) - \phi_i(\boldsymbol{x}_0,k) - F_i(\phi(\boldsymbol{x}_0,k))}{\Delta x} - \delta_{ij}$$
$$= \left.\frac{\partial F_i}{\partial x_j}\right|_k . \tag{16}$$

To derive a quantity analogous to Eq. (16) for the ODE framework in a comparable way, we use the new definition in Eq. (14′) and the assumption in Eq. (12). We can then easily derive the following metric of a parametric interaction strength, which we define as the *instantaneous* interaction strength (IIS)($\mathbf{S}^{I,adj}$):

$$S_{i,j}^{I,adj}(k\tau) \equiv \tau \left.\frac{\partial f_i}{\partial x_j}\right|_{t=k\tau} . \tag{17}$$

Note that when evaluating the accuracy of the S-map (or any equation-free) methods in inferring the interaction Jacobian or interaction strengths through the simulated data from the dynamical systems, it is important to distinguish the sources of the model (DE or ODE). The metric in Eq. (16) is mathematically equivalent to the interaction Jacobian (with the adjustment, $\mathbf{J} = \mathbf{S}^{adj} + \mathbf{I}_n$) for the data generated from a DE system while the metric in Eq. (17) is just an approximation of the interaction Jacobian (with the adjustment, $\mathbf{J} \approx \mathbf{S}^{I,adj} + \mathbf{I}_n$) for the data generated from an ODE system (**Table 2**). It is worth noting that recent studies (Kawatsu, 2024; Munch et al., 2020) already recognized the importance of distinguishing the source of parametric models and the gap between **J** and $\mathbf{S}^{adj}$, proposing an alternative parametric Jacobian for ODE systems as $[\exp\,(\mathbf{S}^{I,adj})]_{i.j}$. In fact, this metric can be approximated as Eq. (13) when $\tau$ is small.

To highlight the contrast with the instantaneous interaction strength defined in Eq. (17), we need to explicitly derive another parametric interaction strength based on the definition in Eq. (14′) without using the assumption in Eq. (12). We define this metric as the *cumulative* interaction strength (CIS)($\mathbf{S}^{C,adj}$) with a time interval $\tau$:

$$S_{i,j}^{C,adj}(k\tau) \equiv S_{i,j}^{adj}(k\tau) \underset{\text{Eq.(14')}}{=} \lim_{\Delta x \to 0} \frac{\varphi_i(\varphi(\boldsymbol{x}_0, k\tau) + \Delta \boldsymbol{x}_j, \tau) - \varphi_i(\varphi(\boldsymbol{x}_0, k\tau), \tau)}{\Delta x} - \delta_{ij}. \tag{18}$$

This can be further generalized through distinguishing the time interval and time scale as shown in **Appendix 2**. Here, the first term of the r.h.s of Eq. (18) represents the alternative parametric Jacobian ($\mathbf{S}^{C,adj} + \mathbf{I}_n$) that considers cumulative effects of time-continuous changes occurring in the interval $\tau$. In order to effectively calculate the numerator of the differential quotient, the differences between the perturbed orbit and the unperturbed orbit starting at $t = k\tau$ is defined as:

$$\boldsymbol{\gamma}(T) \equiv \varphi(\varphi(\boldsymbol{x}_0, k\tau) + \Delta \boldsymbol{x}_j, T) - \varphi(\varphi(\boldsymbol{x}_0, k\tau), T), 0 < T < \tau. \tag{19}$$

While the value of the difference at time $T = \tau$, $\boldsymbol{\gamma}(\tau)$, in Eq. (19) can be obtained by numerically solving the original nonlinear ODEs described in Eq. (4), we propose an alternative method that relies on a simpler ODE.

When $\Delta x$ is infinitesimal, the function $\boldsymbol{\gamma}(T)$ in Eq. (19) can be evaluated as the solution of the following linearized ODE system (without error when $\Delta x \rightarrow 0$), using the Jacobian matrix of the function ***f*** in the ODE system [4] (**Appendix 3.1**):

$$\frac{d\boldsymbol{\gamma}(T)}{dT} = \mathbf{J}^{\boldsymbol{ST}}(T)\boldsymbol{\gamma}(T),\ \ \boldsymbol{\gamma}(0) = \Delta \boldsymbol{x}_j. \tag{20}$$

Note that this is a linear ODE but the state transition matrix $\mathbf{J}^{ST}$ is time-dependent and should be evaluated at each time *T* with the *unperturbed* orbit $\varphi(\varphi(\boldsymbol{x}_0, k\tau), T)$. The numerical method is briefly summarized at **Appendix 3.2**. Also note that the state transition matrix $\mathbf{J}^{ST}$ in Eq. (20) is not the interaction Jacobian but a quantity analogous to the instantaneous interaction strengths without the parameter $\tau$ (i.e., $J_{i,j}^{ST}(T) = \frac{\partial f_i}{\partial x_j}\Big|_T$ and compare it with Eq. (17)).

Since such a time-dependent (non-autonomous) ODE (Eq. (20)) cannot be solved analytically in general, we need to solve it numerically. When we obtain the numerical solution of $\gamma(T)$ at $T = \tau$ (denoting as $\tilde{\gamma}(\tau)$) with a very small $\Delta x$, either from directly solving Eq. (19) with the nonlinear ODE in Eq. (4) or solving the linearized ODE in Eq. (20), the cumulative interaction strengths are given by:

$$S_{i,j}^{C,adj}(k\tau) \approx \frac{[\tilde{\gamma}(\tau)]_i}{\Delta x} - \delta_{ij}. \tag{21}$$

Although Eq. (21) is presented as the approximation ($\approx$), it only includes the numerical errors of $\tilde{\gamma}(\tau)$ from $\gamma(\tau)$ and those due to $\Delta x$ being not infinitesimal. In contrast, using the instantaneous interaction strength (Eq. (17)) is intrinsically an approximation through neglecting the temporal changes of the function $f_i$. Since we are often interested in a dynamical system that has chaotic attractors with initial value sensitivity, $\gamma(T)$ may show the exponential growth over time. Consequently, it is unclear how large $\tau$ can be used when numerically calculating CIS with Eqn. (21) (see **Appendix 4**). In the next section, we will demonstrate that the CIS metric given in Eq. (21) is numerically identical when evaluated using two different methods: direct evaluation of the nonlinear ODE and evaluation based on the linearized ODE.

## 5. Applications to parametric DE and ODE models

Unlike previous research, except for Ushio et al. (2018) (**Table 1**), this study does not include noise in the time series generation process because our goal is not to evaluate various S-map algorithms in a practical setting. Rather, in the absence of noise-related uncertainties, we concentrate on the performance of the proposed two solutions.

### *5.1. Application to 4 species host-parasitoid DE model*

As an example for the DE framework, we used the two host species with two parasitoid species model modified from Chang et al. (2021). The lifecycle of the host species $H_i$ ($i$ = 1 or 2) follows (1) the density-dependent mortality with a rate $a_iH_i$, (2) parasitism by the parasitoid species $P_j$ ($j$ = 1 or 2) with a rate $c_{ji}P_j$, (3) individuals escaping density-dependent mortality and parasitism reproduce $r_i$ individuals contributing to the next generation, and (4) density-independent (background) mortality with a rate $m_{H,i}$. The survived individuals also contribute to the next generation. In the lifecycle of the parasitoid species $P_j$, the parasitism on the hosts first occurs. Repeated parasitism on a single host individual is considered but the single, earliest parasitism succeeds in the maturation and contributes to the next generation (see Chang et al. 2021 for the derivation of its mathematical function). The final stage of the parasitoid lifecycle is the density-independent (background) mortality with a rate $m_{P,j}$, following that the survived individuals also contribute to the next generation. The parameters $m_{H,i}$ and $m_{P,j}$ characterize the overlapping-generation and the model converges to the non-overlapping generation, multispecies Nicholson-Bailey model (Chang et al., 2021) when $m_{H,i}$ and $m_{P,j}$ >> 1. The equations are given by:

$$
\begin{aligned}
H_{i,k+1} = H_{i,k} + & \underbrace{r_i exp[-a_i H_{i,k} - c_{1i}P_{1,k} - c_{2i}P_{2,i}]H_{i,k}}_{\text{birth from hosts that escape intraspecific competition and parasitism}} \\
& - \underbrace{\left(1 - exp[-a_i H_{i,k} - c_{1i}P_{1,k} - c_{2i}P_{2,k} - m_{H,i}]\right)}_{\substack{\text{mortality by intraspecific competition,} \\ \text{parasitism, and natural mortality}}} H_{i,k} \\
= H_{i,k} + & F_{H,i}(H_{1,k}, H_{2,k}, P_{1,k}, P_{2,k}, k), \quad (i = 1{,}2) \qquad (22) \\
P_{j,k+1} = P_{j,k} + & \sum_{i=1,2} \underbrace{exp(-a_i H_{i,k})H_{i,k}}_{\text{availability of hosts}} \underbrace{\left(1 - exp[-c_{1i}P_{1,k} - c_{2i}P_{2,k}]\right)}_{\text{probability of parasitism}} \underbrace{\frac{c_{ji}P_{j,k}}{c_{1i}P_{1,k} + c_{2i}P_{2,k}}}_{\substack{\text{probability with which} \\ \text{parasitim by } P_j \\ \text{is earlier than that by } P_{j'}}} \\
& - \underbrace{\left(1 - exp(-m_{P,j})\right)}_{\text{background mortality}} P_{j,k} \quad (j = 1{,}2) \\
= P_{j,k} + & F_{P,j}(H_{1,k}, H_{2,k}, P_{1,k}, P_{2,k}, k). \quad (j = 1{,}2) \qquad (23)
\end{aligned}
$$

Excluding the transient phase long enough ($k$ = 0 to 1800) with the initial condition ($P_1$, $P_2$, $H_1$, $H_2$) = (1.0, 0.5, 0.1, 0.2), we used the state variables from $k$ = 1801 to 2000 (i.e., 200 data points) for the multivariate S-map (Deyle et al., 2016) and MDR S-map (Chang et al., 2021). Since the multivariate S-map is the basis of other S-map variants (**Table 1**), we denote the former as *standard*

S-map hereafter. When applying these two S-map algorithms, the raw data was standardized to the data with mean 0 and SD = 1. For the theoretical benchmark $\mathbf{S}_{i,j}^{adj}(k)$ in Eq. (16), we used the part of the partial derivative coefficients (one diagonal and one off-diagonal element) given as:

$$\frac{\partial F_{H,1}}{\partial H_{1,k}} = -1 + (1 - a_1 H_{1,k})(r_1 + e^{-m_{H,1}})e^{-a_1 H_{1,k} - c_{11} P_{1,k} - c_{21} P_{2,k}}, \quad (24)$$

$$\frac{\partial F_{H,1}}{\partial P_{1,k}} = c_{11}(-r_1 - e^{-m_{H,1}})H_{1,k} e^{-a_1 H_{1,k} - c_{11} P_{1,k} - c_{21} P_{2,k}}, \quad (25)$$

while we do not need to consider the problem 2 and its solution in the DE framework. Corresponding to the standardization for the S-map methods, we needed to normalize the parametric values of the off-diagonal element (*i*, *j*), by the ratio of standard deviations $\sigma_j/\sigma_i$. More specifically, $\frac{\partial F_{H,1}}{\partial P_{1,k}}$ was normalized by $\sigma_{P_1}/\sigma_{H_1}$.

The adjustment of S-map coefficients by $\delta_{ij}$ worked well, indicating that the magnitude of the intraspecific effect of $H_1$ on $H_1$ is comparable to interspecific ones (from $P_1$ to $H_1$ and from $P_2$ to $H_2$)(**Fig. 1**). If the diagonal element was not adjusted (black point in **Fig. 1**), S-map coefficients would give the impression that intraspecific effect is always positive, fluctuating around 1.0, and its absolute magnitude is greater than interspecific effects, which is against biological intuition. Note that the adjustment of S-map coefficients by $\delta_{ij}$ in Eq. (15) modifies only the intra-specific elements but not inter-specific elements. Also, note that the effect of $H_2$ on $H_1$ is theoretically zero ($\frac{\partial F_{H,1}}{\partial H_{2,k}} = 0$); indeed, this is effectively estimated by the S-map coefficient (from $H_2$ to $H_1$).

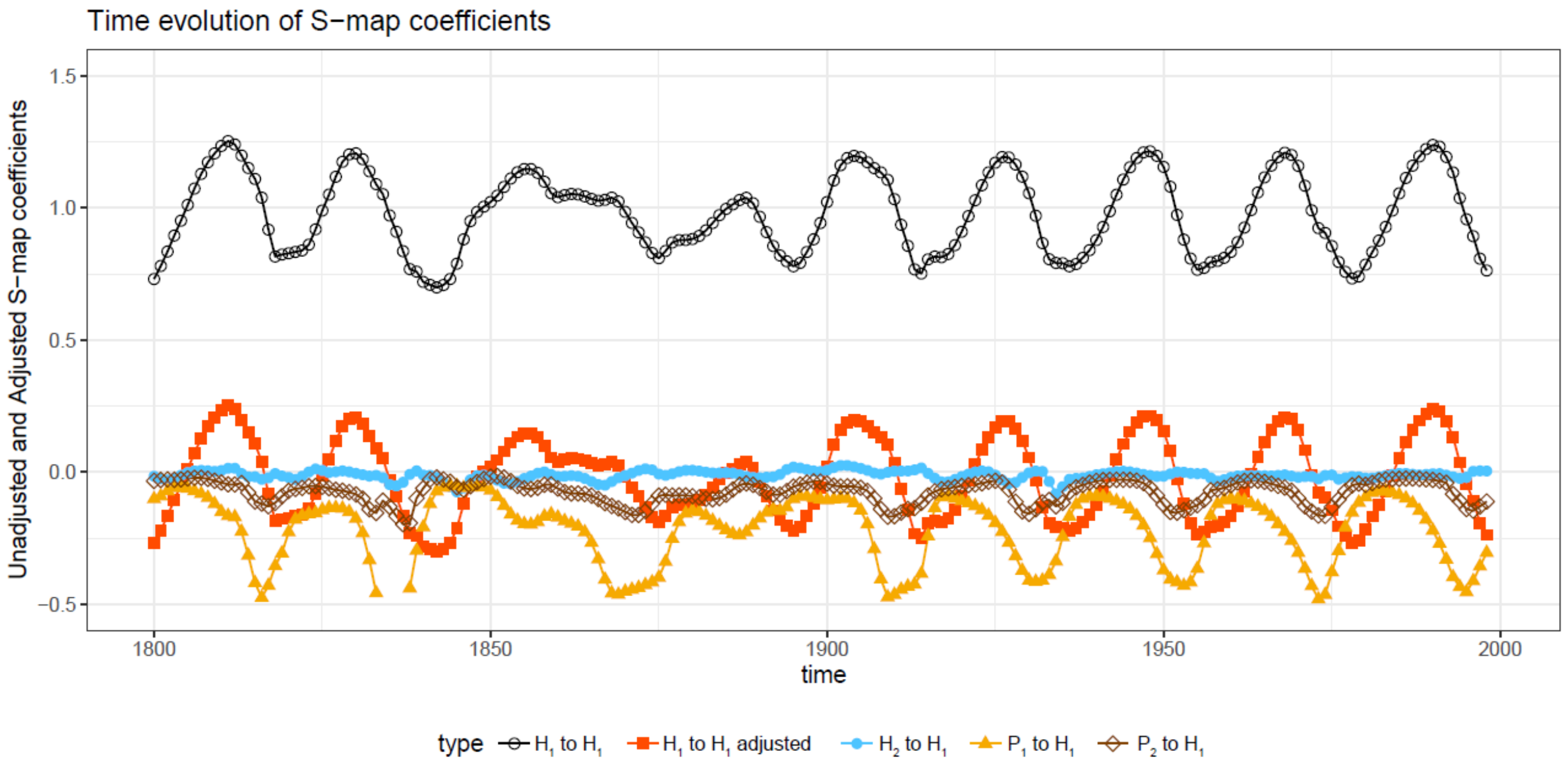

**Figure 1 Diagonal and off-diagonal elements inferred from standard S-map method.** Time evolution of the S-map coefficient from $H_1$ to $H_1$ that were not adjusted (black line with circle marks), adjusted S-map coefficient from $H_1$ to $H_1$ (red line with rectangular marks), from $H_2$ to $H_1$ (light blue line with circle marks), from $P_1$ to $H_1$ (orange line with triangular marks), and from $P_2$ to $H_1$ (brown line with diamond marks). The interaction strength from $H_2$ to $H_1$ is theoretically always zero. The parameter values: $r_1 = r_2 = 2.0$, $a_1 = a_2 = 0.1$, $c_{11} = c_{22} = 0.3$, $c_{21} = c_{12} = 0.1$, $m_{H1} = m_{H2} = 0.1$, and $m_{P1} = m_{P2} = 0.1$. In this example, the off-diagonal element of the theoretical interaction strength was normalized by $\sigma_{P_1}/\sigma_{H_1}$ = 0.477/0.242.

In this example, with comparing to the parametric interaction strength $\mathbf{S}^{adj}(k)$ in Eq. (16) as the benchmark, the standard S-map coefficients for the diagonal element predicted better than MDR S-map coefficients (**Fig. 2ab**). The correlation coefficients with parametric interaction strength were 0.978 and 0.671 for the standard and MDR S-map coefficients, respectively. Similarly, for the off-diagonal element, the gaps between the estimate and the benchmark tended to be greater with the MDR S-map especially at local minimum than with the standard S-map (**Fig. 2cd**). The correlation coefficients with parametric interaction strength were 0.962 and 0.971 for the standard and MDR S-map coefficients, respectively.

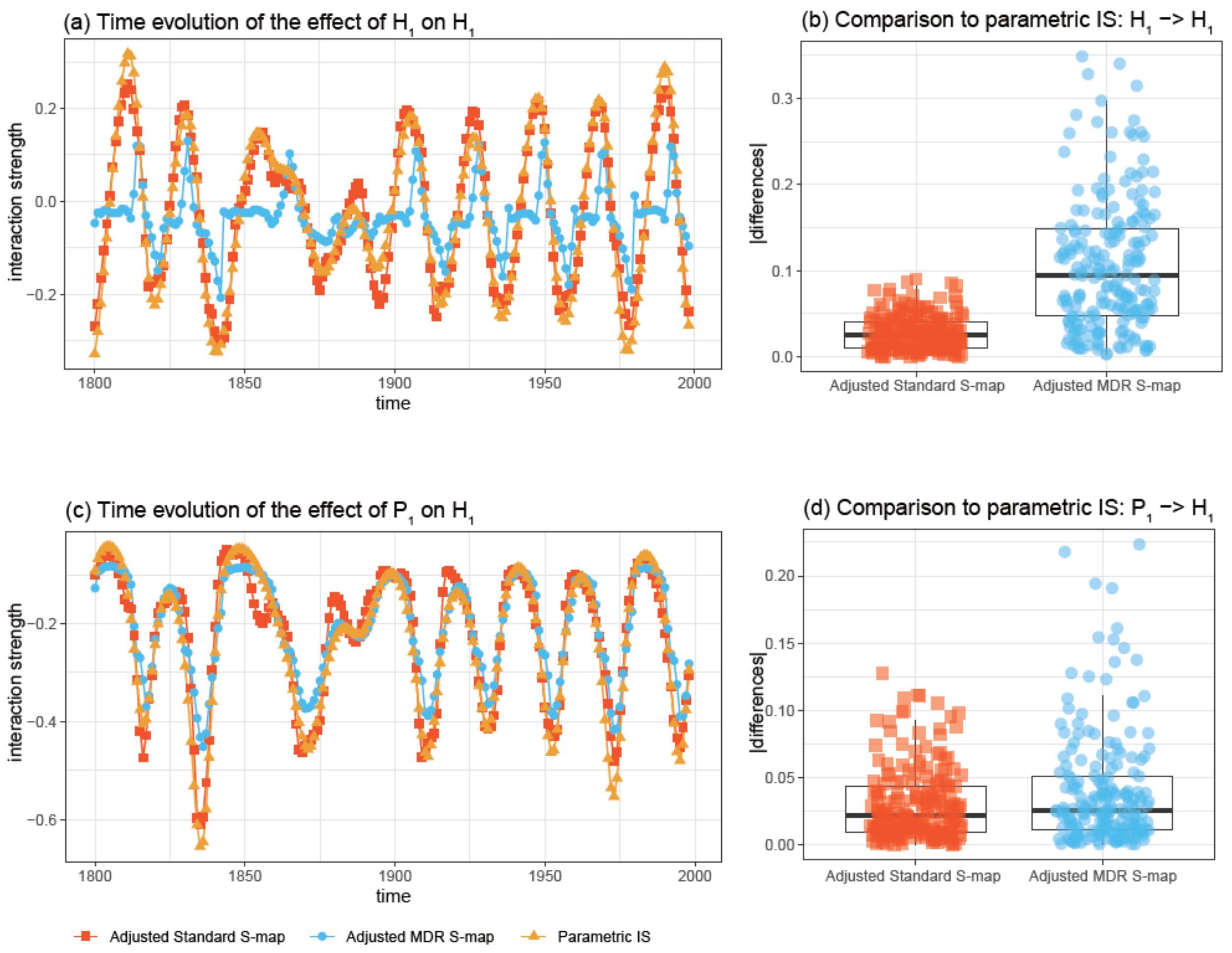

**Figure 2 Comparison of diagonal and off-diagonal elements for DE system.** (a)(c) Time evolution of interaction strengths estimated from the adjusted standard S-map coefficient (red, rectangle), the adjusted MDR S-map coefficient (light blue, circle), and the parametric interaction strength (orange, triangle) for the diagonal element (**a:** from $H_1$ to $H_1$) and the off-diagonal element (**c**: from $P_1$ to $H_1$). (b)(d) Distribution of the absolute value of the differences between S-map coefficients (standard S-map: red, rectangle, MDR S-map: light blue, circle) and parametric interaction strengths for the diagonal element (**b:** from $H_1$ to $H_1$) and the off-diagonal element (**d**: from $P_1$ to $H_1$). In this example, the off-diagonal element of the parametric interaction strength was normalized by $\sigma_{P_1}/\sigma_{H_1} = 0.477/0.242$.

### *5.2. Application to 5 species coupled food chain ODE model*

As an example for the ODE framework, we used the five species coupled food chain model (Deyle et al., 2016), which consists of a resource species ($R$), two consumer species ($C_1$, $C_2$) sharing $R$, and two specialist predator species ($P_1$, $P_2$) feeding on $C_1$ and $C_2$, respectively. The model equations are given by:

$$
\begin{aligned}
\frac{dP_i}{dt} &= v_i\lambda_i\frac{P_iC_i}{C_i+C_i^*} - v_iP_i \equiv f_{P_i}, \quad (i=1{,}2)\\
\frac{dC_i}{dt} &= \mu_i\kappa_i\frac{C_1R}{R+R^*} - v_i\lambda_i\frac{P_iC_i}{C_i+C_i^*} - \mu_iC_i \equiv f_{C_i}, \quad (i=1{,}2)\\
\frac{dR}{dt} &= R\left(1-\frac{R}{k}\right) - \sum_{i=1,2}\mu_i\kappa_i\frac{C_iR}{R+R^*} \equiv f_R.
\end{aligned}
\tag{26}
$$

The default setting for obtaining the discrete-time data is $\tau = 5$ as the same as in Deyle et al (2016). We numerically solve the system using the fourth-order Runge-Kutta method with a fixed interval $\Delta t = 0.01$ from $t = 0$ to 2000 with the initial condition ($P_1$, $P_2$, $C_1$, $C_2$, $R$) = (0.7, 0.8, 0.5, 0.8, 1.0). Excluding the transient phase values ($t = 0$ to 999), we used the state variables from $t = 1000$ to $t = 1999$ (i.e., 200 data points with the interval $\tau = 5$) for the standard S-map (Deyle et al., 2016) and MDR S-map (Chang et al., 2021). For the cumulative interaction strength (CIS), we used the list of partial derivatives (**Appendix 5**) for the element of the state transition matrix $\mathbf{J}^{ST}$ in Eq. (20). As is described just below Eq. (20), $\mathbf{J}^{ST}$ should be evaluated at each moment of time $t$ with the unperturbed (numerical) solution of the model in Eq. (26)(see also **Appendix 3.2** for the numerical method).

We compared the adjusted S-map coefficients (Standard and MDR S-map algorithms) defined in Eq. (15), instantaneous interaction strength (IIS) in Eq. (17), and cumulative interaction strength (CIS) in Eq. (18). When applying S-map methods, the raw data was standardized to the data with mean 0 and SD = 1. Corresponding to this standardization, we needed to normalize the values of the off-diagonal element ($i,j$), by the ratio of standard deviations $\sigma_j/\sigma_i$, of the parametric interaction strengths in Eqs. (17) and (18). The solution 1 for the S-map coefficients (adjustment by $\delta_{ij}$) worked well for the diagonal element; the intraspecific effect of $C_1$ on $C_1$ can be positive or negative depending on the system state and time (**Figure 3a**).

It is also found that the solution 2 is important; there was a substantial gap between two parametric Jacobians and consequently two parametric interaction strengths, IIS and CIS evaluated by Eq.(21). In particular, IIS overestimated the magnitude of local minimum values while it underestimated that of local maximum values (**Figure 3a**). In fact, the gaps between IIS and CIS directly evaluated by Eq. (19) were greater than those between S-map coefficients and CIS (**Figure 3b**). The correlation between IIS and CIS was weak ($\rho = 0.706$). Also note that the CIS evaluated by the linearized ODE in Eq. (20) (and see **Appendix 5**) and that by the original nonlinear ODE in Eqs.

(19) and (26) are numerically identical (**Figure 3b,** orange triangles). These results regarding IIS and CIS were qualitatively similar in the case of the off-diagonal element (**Figure 3cd**). Also. it was numerically confirmed that the IIS in Eq. (17) converged to CIS in Eq. (18) when $\tau$ is small (e.g. $\tau$ = 0.01 and 0.5, which corresponds to $\tau_2$ in **Appendix 2**). However, the deviations emerged with $\tau$ = 1.0 and became substantial with $\tau \geq 2.0$ (**Figure 4**).

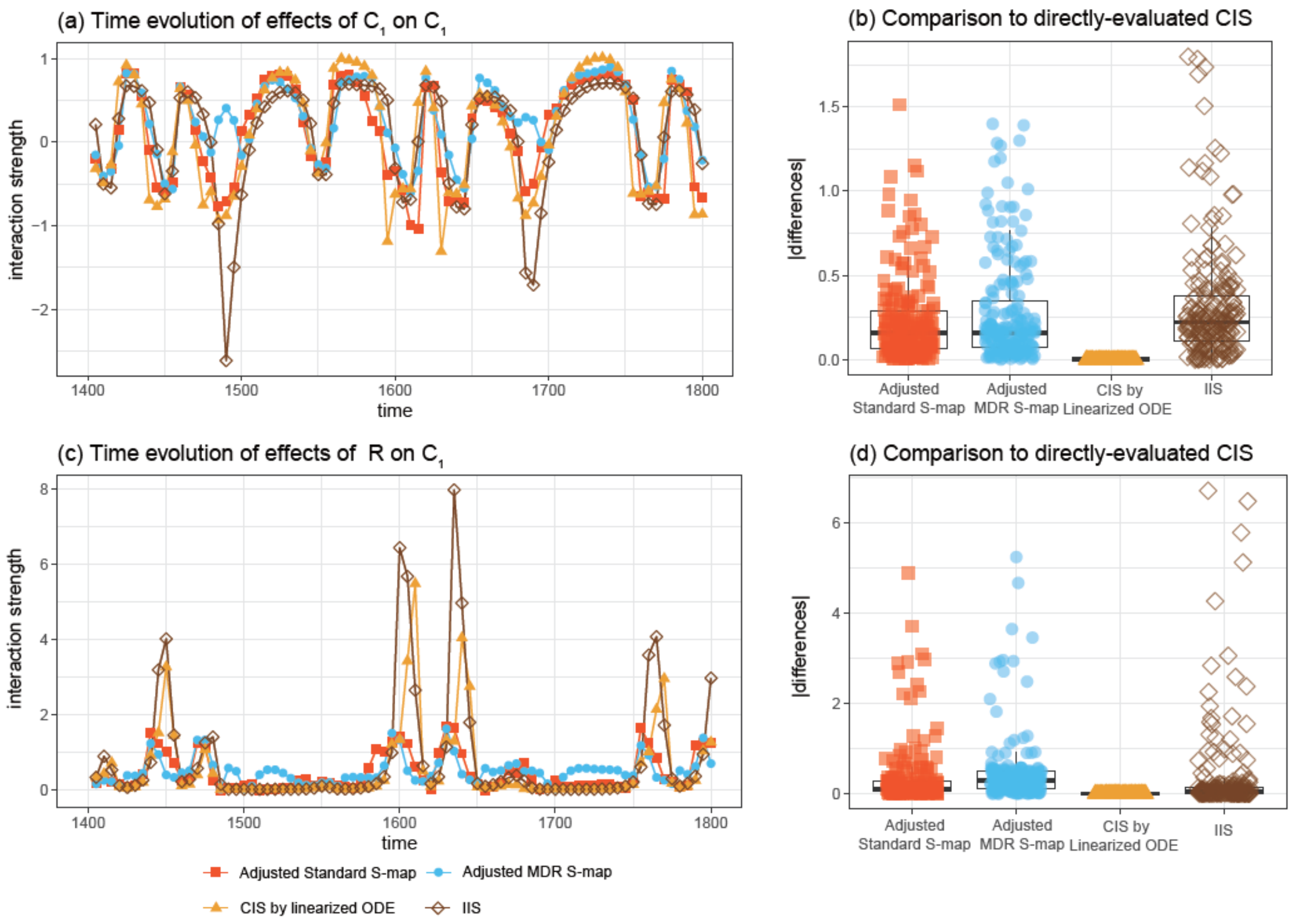

**Figure 3. Comparison of diagonal and off-diagonal elements for ODE system.** (a)(c) Time evolution of interaction strengths estimated from the adjusted standard S-map coefficient (red, rectangle), the adjusted MDR S-map coefficient (light blue, circle), CIS evaluated by the linearized ODE (orange, triangle), and IIS (brown, diamond) for the diagonal element (**a:** from $C_1$ to $C_1$) and the off-diagonal element (**c**: from $R$ to $C_1$). (b)(d) Distribution of the absolute value of the errors of S-map coefficients (standard S-map: red, rectangle, MDR S-map: light blue, circle), CIS evaluated by the linearized ODE (orange, triangle), and IIS (brown, diamond), compared to CIS directly evaluated by the nonlinear ODE, for the diagonal element (**b:** from $C_1$ to $C_1$) and the off-diagonal element (**d**: from $R$ to $C_1$). Parameter values are: $v_1 = 0.1$, $v_2 = 0.07$, $\lambda_1 = 3.2$, $\lambda_2 = 2.9$, $C_1^* = C_2^* = 0.5$, $\mu_1 = \mu_2 = 0.15$, $\kappa_1 = 2.5$, $\kappa_2 = 2.0$, $R^* = 0.3$, and $k = 2$. In this example, the off-diagonal elements of the parametric interaction strength (CIS and IIS) were normalized by $\sigma_R/\sigma_{C_1}$= 0.424 /0.425. The size of perturbation to approximately calculate the CIS in two ways (Eqs. (19) and (20)) was set as $\Delta x = 0.01$.

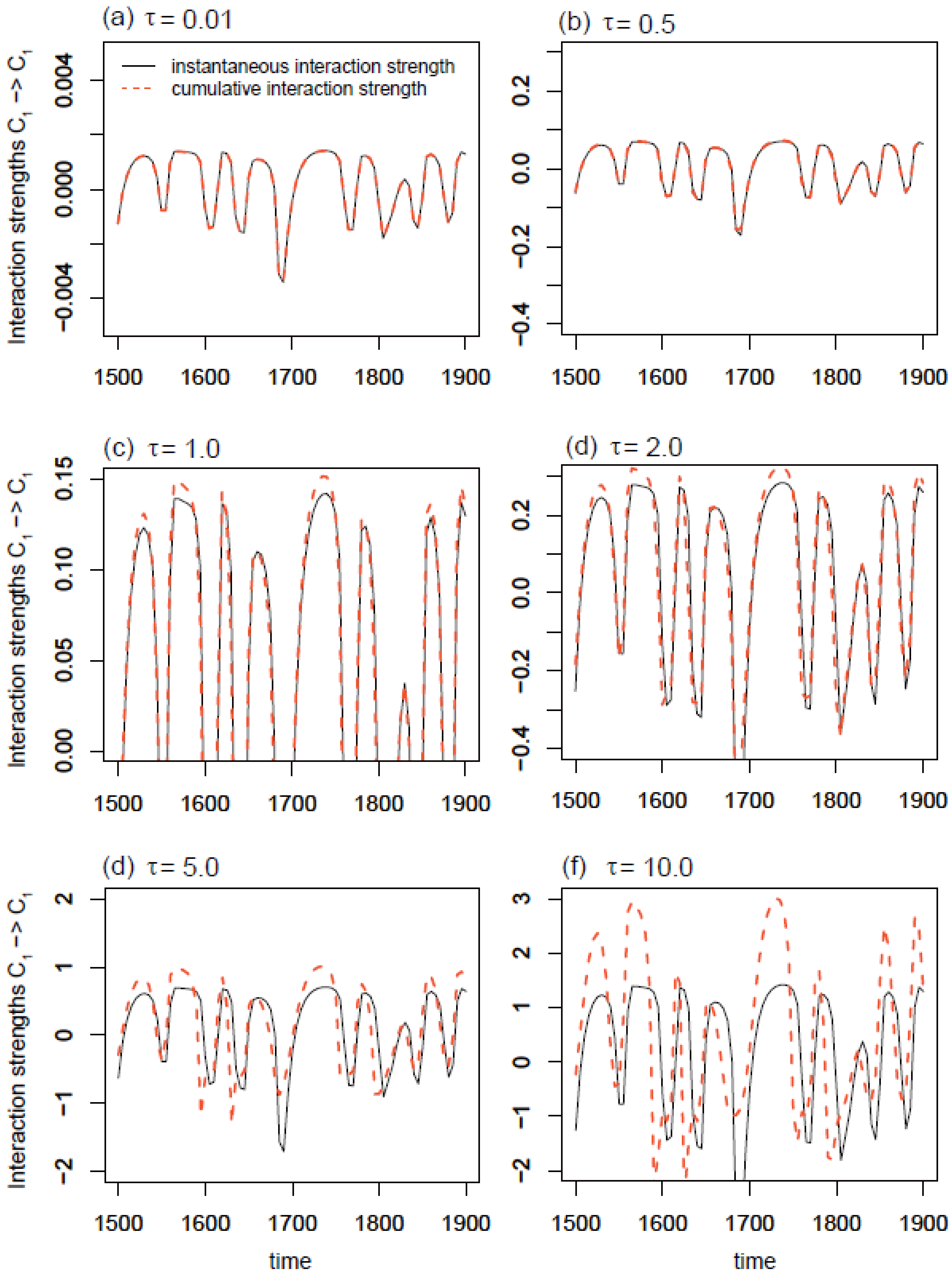

**Figure 4. Comparison of IIS and CIS.** We used the identical data points generated from the ODE with the interval 5.0 ($\tau_1$ = 5.0, see **Appendix 2**) for all the results. However, we changed the time scale for evaluating

interaction strength ($\tau_2$, see **Appendix 2**) in the range between 0.01 and 10.0 for both IIS and CIS. Such separation of time interval between data points and time scale for evaluating interaction strengths can be also applied to the DE framework (**Appendix 2**). The parameter setting was identical to the results shown in **Figure 3**.

## 6. Discussion

### 6.1. Key Findings: Identifying Problems and Proposing Solutions

Our mathematical reintroduction of the interaction Jacobian in Eqs. (6a) and (7) clarified previously ambiguous and inconsistent biological interpretations of the interaction Jacobian, thereby identifying and resolving two potential problems in applying S-map methods within EDM.

The first problem was the discrepancy between mathematical definition of the interaction Jacobian and its biological interpretation as interaction strength. For better consistency with biological intuition, we proposed an adjustment to the interaction strength particularly for the diagonal elements. Our definition of the adjusted interaction strength from variable $x_j$ to $x_i$ is the *subsequent* marginal change of the recipient variable $x_i$ at the time step $k + 1$, occurring after a unit-sized perturbation on the donor variable $x_j$ at the time step $k$ (see **Fig. 5** for graphical presentation). For the off-diagonal elements ($j \neq i$), no change is imposed on the recipient variable $x_i$ at the time step $k$; that is, the *initial* marginal change is zero. Therefore, the subsequent marginal change is identical to the *total* marginal change, because the total marginal change equals the initial marginal change (0) plus subsequent marginal change. As a result, the adjusted interaction strength remains identical to the value from the original definition. Conversely, for the diagonal elements ($j = i$), because a unit-sized change is imposed on the recipient variable $x_i$ itself at the time step $k$, the initial marginal change is 1 resulting from the unit-sized perturbation. Consequently, the subsequent marginal change must be calculated by subtracting 1 from the total marginal change. In the absence of subsequent changes after the unit-sized perturbation at the time step $k$, the total marginal change equals 1, and the adjusted interaction strength of $x_i$ on itself becomes zero.

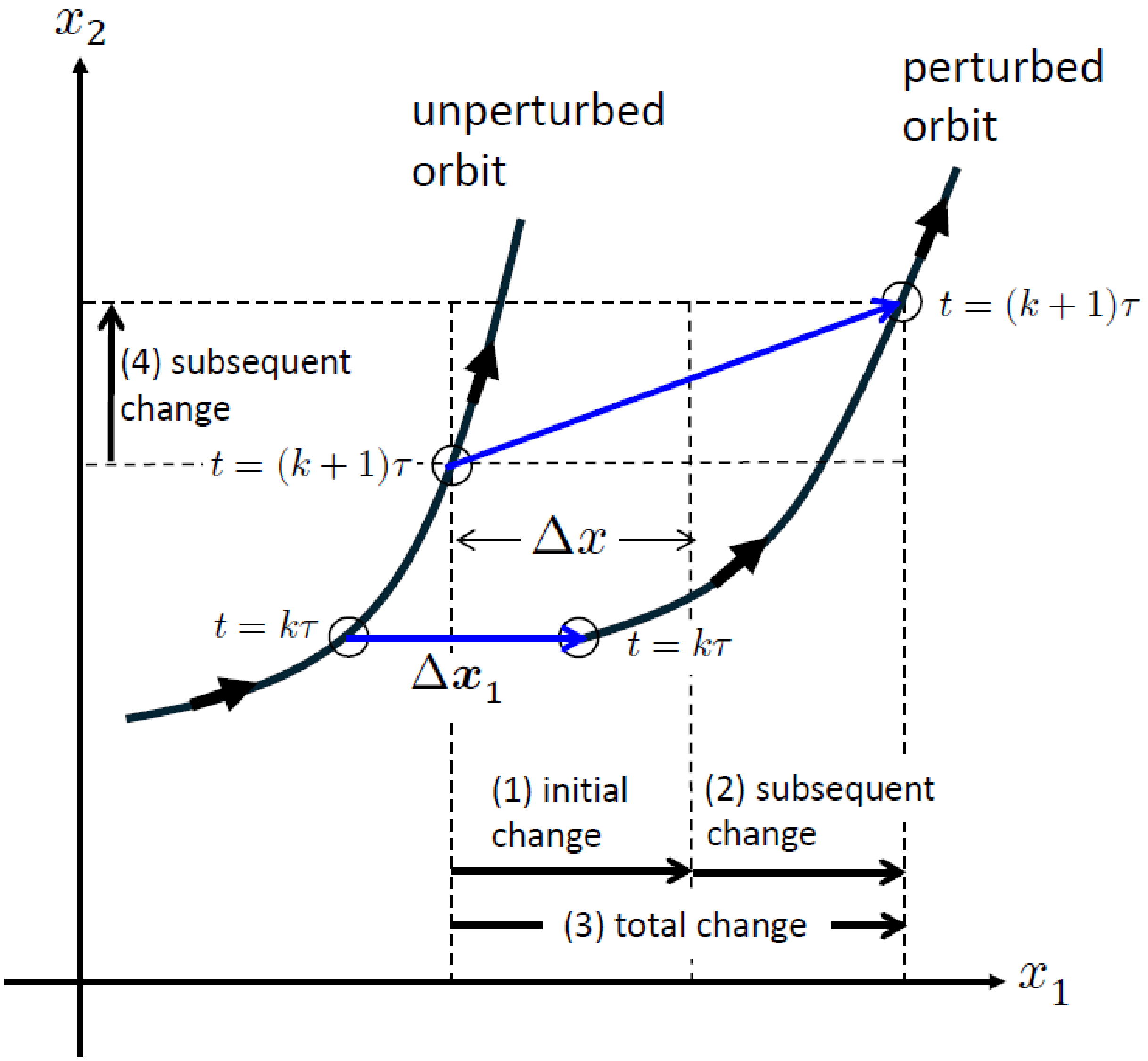

**Figure 5 Relationship between interaction Jacobian and interaction strengths**. Using a hypothetical example focusing on two elements only ($x_1$ and $x_2$), we illustrate the relationships between the interaction Jacobian ($J_{1,1}$ and $J_{2,1}$), interspecific interaction strengths (the effect of *sp*. 1 *on* sp. 2), and intraspecific interaction strengths (the effect of *sp*. 1 on *sp*. 1). When adding the perturbation given by the vector $\Delta \boldsymbol{x}_1$ to the unperturbed orbit at time $t = k\tau$, we can quantify the differences between the unperturbed and perturbed orbits, as the consequence of this perturbation, at $t = (k+1)\tau$, for both the diagonal component ($x_1$) and the off-diagonal element ($x_2$). For the diagonal element, (1) and (2) correspond to the *initial* change and *subsequent* change, respectively, following that the *total* change (3) equals (1) + (2). For the off-diagonal element, the *initial* change is zero and (4) corresponds to the *subsequent* change, following that the *total* change equals (4). By dividing these quantities by $\Delta x$ and taking the limit as $\Delta x \to 0$, we can easily find that the quantifies from

(3) and (4) become equal to the interaction Jacobian $J_{1,1}$, and $J_{2,1}$, respectively, which are also denoted as the total *marginal* changes. Similarly, the quantities from (2) and (4) become equal to the adjusted intraspecific interaction strength $S_{1,1}^{adj}(k\tau)$ and the adjusted interspecific interaction strength $S_{2,1}^{adj}(k\tau)$, respectively, which are also denoted as the subsequent *marginal* changes. Although this graphical example illustrates to a case within the ODE framework, the relationships between the interaction Jacobian and the adjusted interaction strengths remain consistent within the DE framework.

The second problem was a mathematical discrepancy between the interaction Jacobian and the parametric Jacobian, which arises exclusively within the ODE framework. To address this discrepancy, we proposed a new theoretical benchmark for simulated data generated by an ODE model. The existing parametric Jacobian from an ODE model only accounts for the instantaneous values of the Jacobian coefficients of the function ***f*** in the r.h.s. of the ODE at the focal time point. This overlooks the continuous changes in the state variables over the finite interval between the focal time point and the next. The cumulative interaction strength (CIS) with the modified parametric Jacobian ($\mathbf{S}^{C,adj} = \mathbf{J} - \mathbf{I}_n$), serves as a more accurate theoretical benchmark by explicitly accounting for the cumulative effects of temporally varying state variables within a given interval. While the instantaneous interaction strength (IIS) provides a reasonable approximation of the interaction Jacobian ($\mathbf{J} \approx \mathbf{S}^{I,adj} + \mathbf{I}_n$) when the time interval is sufficiently short, CIS is recommended for most intervals considered in recent studies (e.g., Deyle *et al.* 2016). Munch, Rogers and Sugihara (2023) pointed out the potential discrepancy between IIS and empirically inferred S-map coefficients over large time intervals, which our study indirectly confirmed through a comparison between CIS (serving as an appropriate benchmark for S-map coefficients) and IIS.

## 6.2 Implications for Ecological Applications: Addressing Overestimation in Intraspecific Effects (Problem 1) with a Simple Adjustment (Solution 1)

In our revised definition of interaction strength in Eqs. (16) - (18), the effect of a variable $x_i$ on itself is directly linked to the partial derivatives of the population growth rate function, denoted as ***F*** for the DE framework in Eq. (3) and ***f*** for the ODE framework in Eq. (4). This fits well with the biological intuition regarding intraspecific effects. Therefore, in non-parametric approaches using observed time series, this adjusted definition should replace the diagonal element of the interaction Jacobian, such as those inferred from S-map methods. Without this adjustment, overestimation can occur, as the diagonal element of the interaction Jacobian and corresponding S-map coefficients always includes the initial marginal change (= 1), thereby apparently indicating positive intraspecific effects.

However, two important cautions must be considered. First, as demonstrated by Song and Saavedra (2021), intraspecific effects defined by the interaction Jacobian do not solely represent the direct effect of intraspecific density-dependence. Instead, they reflect the aggregated effects of all possible interaction pathways involving the focal species. Therefore, the intraspecific interaction strength inferred from any relevant methods in **Table 1** with the adjusted linear coefficient in Eq. (15) should not be mistaken for the intraspecific density-dependent coefficient used in parametric population dynamics model, such as generalized Lotka-Volterra models, nor is it proportional to it (see Song and Saavedra, 2021 for a more mathematically explicit argument). For instance, it would be inappropriate to calculate the ratio of intra- and interspecific interaction strengths inferred by the adjusted linear coefficients in Eq. (15) as an index for species coexistence. Although this ratio serves as a good index in the Lotka-Volterra model framework, it does not hold the same meaning in the context of non-parametric S-map methods, especially because many of S-map methods do not rely on any parametric model equations to describe the system's governing dynamics.

Second, when considering any given perturbation ($\Delta \boldsymbol{p} \in \mathbb{R}^n$) on system state vector, the interaction Jacobian **J** itself acts as the transition matrix for the temporal propagation of the initial perturbation at the focal time point. This is expressed as $\Delta \boldsymbol{p}' = \mathbf{J} \Delta \boldsymbol{p} + O(\|\Delta \boldsymbol{p}\|^2)$ where $\Delta \boldsymbol{p}'$ represents the perturbed vector in the next time step. Therefore, when focusing on dynamical stability, in terms of the divergence of the system state against perturbation on system state variable, with a method such as dynamic eigenvalue (DEV, Grziwotz et al., 2023), it is essential to use the interaction Jacobian matrix itself rather than the matrix of adjusted interaction strengths.

## 6.3 Implications for Method Development: Uncovering Hidden Approximations in Previous Theoretical Benchmarks (Problem 2) and Proposing Mathematical Improvement (Solution 2)

When one develops a new non-parametric method, such as a modified S-map algorithm, to infer the interaction Jacobian, it is common to evaluate its performance using the simulated data generated from the parametric dynamical models (either DEs or ODEs) and comparing the inference with the parametric Jacobian. When ODEs are used for such evaluation, the cumulative interaction strength (CIS), which is the modified parametric Jacobian minus $\delta_{ij}$, is a more accurate theoretical benchmark than IIS. Our simple exercise demonstrated that the error size of IIS was the same as or even larger than that of the inference from S-map methods when using the CIS values as correct answers (**Fig. 3bd**). Once the non-parametric method is theoretically evaluated, it is expected to be applied to empirical time series with long time intervals. Consequently, IIS should generally not be used for theoretical evaluations, except in cases where empirical time series data with very short time intervals are likely available (**Fig. 4**). Beyond S-map methods, the interaction Jacobians and interaction strengths can be explored through other EDM approaches, such as Gaussian process methods (Johnson and Munch, 2022; Tsai et al., 2024), where our proposed methods could also be applied for evaluation and benchmarking.

We also confirmed that direct evaluation using the nonlinear ODE with the definition in Eq. (19) and the linearized ODE in Eq. (20) are numerically identical. Although the linearized ODE appears mathematically more formal, its computational cost was larger than that of the direct evaluation, because solving the non-autonomous linearized ODE requests a higher time-resolution solution than the nonlinear ODE (**Appendix 3.2**). Therefore, we would recommend using the direction evaluation method for saving computational costs.

### 6.4 Mathematician's guide for EDM: Explicit Presentation of Methods

This study explicitly reintroduced mathematics regarding the interaction Jacobian and the interaction strength. From a mathematical point of view, the problem 1 resulted from the mismatches between the interaction Jacobian, defined as the partial derivatives (Jacobian) of the flow of dynamical systems, and the Jacobians of the right-hand sides of parametric dynamical model equations, that is, the functions representing population growth rate ($\boldsymbol{F}$ and $\boldsymbol{f}$). To our knowledge, this has been explicitly recognized only in two studies (Kawatsu, 2024; Munch et al., 2020). Therefore, one needs to be careful when interested in any properties of $\boldsymbol{F}$ or $\boldsymbol{f}$, rather than the properties of the flow itself.

The multivariate S-map methods, as representatives of non-parametric approaches for inferring the interaction Jacobian (**Table1**), are just a small part of the core components of EDM. Here, we present a short list of recent studies that present mathematical details for mathematicians who are interested in this field. Song and Saavedra (2021) present an excellent summary of general mathematical issues regarding interaction strength. Edwards et al. (2024) provides the details of mathematics regarding one of the core methods of EDM, the simplex projection. Cenci and Saavedra (2019) proposed the usage of the trace of Jacobian of ODEs as an index of structural stability at non-equilibrium while Grziwotz et al. (2023) focused on the dominant eigenvalue of interaction Jacobian as an index of dynamical stability. The relationship between the two indices stems from the interaction Jacobian and S-map methods, which can be made clearer by more rigorous mathematical analysis.

Statistically and mathematically defining the causal-effect relationship between two focal variables remains an ongoing research topic (Osada et al., 2023). Additionally, improving the accuracy of short-term forecasts remains challenging due to observation noise in time series data, and it continues to be an active area of research (Esguerra and Munch, 2024). We would also like to leave as an open proposition regarding the stability of the numerical methods for calculating the interaction strength (**Appendix 4**).

## 6.5. Conclusion

In summary, our study mathematically reintroduced the interaction Jacobian to clarify ambiguities in its biological interpretation within Empirical Dynamic Modeling. We addressed two key problems: the discrepancy between the mathematical definition of interaction Jacobian and its biological intuition as interaction strength, and the inconsistency between the interaction Jacobian and parametric Jacobian in ordinary differential equation frameworks. By proposing an adjusted definition for interaction strength and introducing the cumulative interaction strength as a more accurate theoretical benchmark, we provided practical solutions that can help more appropriate application and development of S-map methods and related non-parametric approaches. These contributions not only improve methodological approaches in ecological time series analysis but also offer valuable insights for future research in ecological networks and system stability analysis.

**Data availability**

No data were collected for this study. Programming codes are available at https://github.com/tksmiki/interaction_Jacobian.

**Acknowledgment**

This work was supported by Grant-in-Aid for Scientific Research (19H05667, 19H00956, and 23H00538) of JSPS (to TM), CWC is supported by the Taiwan Ministry of Education Yushan Fellow Program (MOE-NTU-113V1034-2) and Taiwan National Science and Technology Council (NSTC 113-2621-M-002-005), and PJK is supported by the Taiwan Ministry of Education Yushan Fellow Program (MOE-110-YSFAG-0003-001-P1) and Taiwan National Science and Technology Council (MOST 111-2621-B-002-001-MY3). CHT is supported by the Taiwan National Science and Technology Council (NSTC 113-2611-M-006 -005 -MY2).

**APPENDIX 1.1** Interaction Jacobian with no net dynamics

When the conditions in Eqs. (8) and (9) are satisfied, we will trivially have the followings:

$$\phi_i(\phi(\boldsymbol{x}_0, k) + \Delta\boldsymbol{x}_j, 1) = \phi_i(\boldsymbol{x}_0, k) + [\Delta\boldsymbol{x}_j]_i = \phi_i(\boldsymbol{x}_0, k) + \delta_{ij}\Delta x, \forall j = 1,2,\dots,n. \qquad (A1)$$

and

$$\varphi_i(\varphi(\boldsymbol{x}_0, k\tau) + \Delta\boldsymbol{x}_j, \tau) = \varphi_i(\boldsymbol{x}_0, k\tau) + [\Delta\boldsymbol{x}_j]_i = \varphi_i(\boldsymbol{x}_0, k\tau) + \delta_{ij}\Delta x, \forall j = 1,2,\dots,n. \qquad (A2)$$

Using Eqs. (6a) and (A1) gives:

$$J_{i,j}(k) = \lim_{\Delta x \to 0} \frac{\phi_i(\boldsymbol{x}_0, k) + \delta_{ij}\Delta x - \phi_i(\phi(\boldsymbol{x}_0, k), 1)}{\Delta x} = \lim_{\Delta x \to 0} \frac{\phi_i(\boldsymbol{x}_0, k) + \delta_{ij}\Delta x - \phi_i(\boldsymbol{x}_0, k)}{\Delta x} = \delta_{ij},$$

in the DE framework where $\delta_{ij}$ is the Kronecker delta.

Notably we obtain the same conclusion with the ODE framework from Eqs. (7) and (A2) as:

$$\begin{aligned} J_{i,j}(k\tau) &= \lim_{\Delta x \to 0} \frac{\varphi_i(\boldsymbol{x}_0, k\tau) + \delta_{ij}\Delta x - \varphi_i(\varphi(\boldsymbol{x}_0, k\tau), \tau)}{\Delta x} \\ &= \lim_{\Delta x \to 0} \frac{\varphi_i(\boldsymbol{x}_0, k\tau) + \delta_{ij}\Delta x - \varphi_i(\boldsymbol{x}_0, k\tau)}{\Delta x} \\ &= \delta_{ij} \ . \end{aligned}$$

**APPENDIX 1.2** How to derive the instantaneous interaction strength

Here we start with Eqs. (5″) and (7):

$$\begin{aligned} J_{i,j}(k\tau) &= \lim_{\Delta x \to 0} \frac{1}{\Delta x} \left[ \varphi_i(\varphi(\boldsymbol{x}_0, k\tau) + \Delta \boldsymbol{x}_j, \tau) - \varphi_i(\varphi(\boldsymbol{x}_0, k\tau), \tau) \right] \\ &= \lim_{\Delta x \to 0} \frac{1}{\Delta x} \left[ \varphi_i(\boldsymbol{x}_0, k\tau) + \delta_{ij}\Delta x + \int_0^\tau f_i\left(\varphi(\varphi(\boldsymbol{x}_0, k\tau) + \Delta \boldsymbol{x}_j, s)\right) ds - \varphi_i(\boldsymbol{x}_0, k\tau) \right. \\ &\qquad \left. - \int_0^\tau f_i(\varphi(\varphi(\boldsymbol{x}_0, k\tau), s)) ds \right] \\ &= \lim_{\Delta x \to 0} \frac{\delta_{ij}\Delta x}{\Delta x} + \lim_{\Delta x \to 0} \frac{1}{\Delta x} \left[ \int_0^\tau f_i\left(\varphi(\varphi(\boldsymbol{x}_0, k\tau) + \Delta \boldsymbol{x}_j, s)\right) ds - \int_0^\tau f_i(\varphi(\varphi(\boldsymbol{x}_0, k\tau), s)) ds \right] \\ &= \delta_{ij} + \lim_{\Delta x \to 0} \frac{1}{\Delta x} \left[ \int_0^\tau f_i\left(\varphi(\varphi(\boldsymbol{x}_0, k\tau) + \Delta \boldsymbol{x}_j, s)\right) ds - \int_0^\tau f_i(\varphi(\varphi(\boldsymbol{x}_0, k\tau), s)) ds \right] \qquad (A3). \end{aligned}$$

Then, under the assumption in Eq. (12), the second term of the r.h.s. of Eq. (A3) is approximated as:

$$\lim_{\Delta x \to 0} \frac{1}{\Delta x} \left[ \left[ \left( f_i(\varphi(\varphi(\boldsymbol{x}_0, k\tau), 0)) + \left. \frac{\partial f_i}{\partial x_j} \right|_{t=k\tau} \Delta x + O((\Delta x)^2) \right) s \right]_0^\tau - \left[ f_i(\varphi(\varphi(\boldsymbol{x}_0, k\tau), 0)) s \right]_0^\tau \right]$$

$$= \tau \left. \frac{\partial f_i}{\partial x_j} \right|_{t=k\tau} . \qquad (A4)$$

Therefore, from Eqn. (A3) and (A4), we have Eq. (13).

**Appendix 2**

The *cumulative* interaction strength defined by Eq. (18) can be naturally extended as the *cumulative* interaction strength at time = $k\tau_1$ with a time scale $\tau_2$:

$$S_{i,j,\tau_2}^{C,adj}(k\tau_1) \equiv \lim_{\Delta x \to 0} \frac{\varphi_i(\varphi(\boldsymbol{x}_0, k\tau_1) + \Delta \boldsymbol{x}_j, \tau_2) - \varphi_i(\varphi(\boldsymbol{x}_0, k\tau_1), \tau_2)}{\Delta x} - \delta_{ij} \qquad (18')$$

The same concept can be applied to the instantaneous interaction strength as:

$$\mathbf{S}_{i,j,\tau_2}^{I,adj}(k\tau_1) \equiv \tau_2 \left.\frac{\partial f_i}{\partial x_j}\right|_{t=k\tau_1} . \qquad (17')$$

When the time interval τ is chosen and consider the interaction strength from *k* to $k + \theta\,(\theta > 1)$ in DE framework, we need to consider the iterations regarding the map ***F***. A new benchmark for the DE system with arbitrary time step $\theta$ (i.e., time scale $\tau\theta$):

$$\gamma(\Delta \boldsymbol{x}_j, \theta) \approx \prod_{\boldsymbol{l}=\mathbf{0}}^{\boldsymbol{\theta}-\mathbf{1}} \left( \boldsymbol{I} + \left.\frac{\partial F_i}{\partial x_j}\right|_{\boldsymbol{x}=\boldsymbol{\varphi}_{\boldsymbol{UP}}(l)} \right) \Delta \boldsymbol{x}_j .$$

**APPENDIX 3.1** How to derive the mathematical formulation in Eq. (20)

Consider the unperturbed orbit and perturbed orbit (at $t = k\tau$) as:

$$\boldsymbol{\theta}_{\boldsymbol{UP}}(T) \equiv \varphi(\varphi(\boldsymbol{x}_0, k\tau), T), \text{and } \boldsymbol{\theta}_{\boldsymbol{P}}(\Delta \boldsymbol{x}_j, T) \equiv \varphi(\varphi(\boldsymbol{x}_0, k\tau) + \Delta \boldsymbol{x}_j, T) \qquad (A5)$$

, respectively.

Then, the difference between the two orbits (Eq.19) is given by:

$$\boldsymbol{\gamma}(\Delta \boldsymbol{x}_j, T) \equiv \boldsymbol{\theta}_{\boldsymbol{P}}(\Delta \boldsymbol{x}_j, T) - \boldsymbol{\theta}_{\boldsymbol{UP}}(T). \qquad (A6)$$

The *i*-th element of the unperturbed orbit is given with $T_1$, $\Delta t > 0$:

$$\theta_{UP,i}(T_1 + \Delta t) = \theta_{UP,i}(T_1) + \int_{T_1}^{T_1 + \Delta t} f_i(\boldsymbol{\theta}_{\boldsymbol{UP}}(s)) ds . \qquad (A7)$$

The $i$-th element of the perturbed orbit is given with $T_1$, $\Delta t > 0$:

$$\theta_{P,i}(\Delta \boldsymbol{x}_j, T_1 + \Delta t) = \theta_{P,i}(\Delta \boldsymbol{x}_j, T_1) + \int_{T_1}^{T_1+\Delta t} f_i\left(\boldsymbol{\theta_P}(\Delta \boldsymbol{x}_j, s)\right) ds$$

$$\underset{A6}{=} \theta_{UP,i}(T_1) + \gamma_i(\Delta \boldsymbol{x}_j, T_1) + \int_{T_1}^{T_1+\Delta t} f_i\left(\boldsymbol{\theta_{UP}}(s) + \boldsymbol{\gamma}(\Delta \boldsymbol{x}_j, s)\right) ds$$

$$= \theta_{UP,i}(T_1) + \gamma_i(\Delta \boldsymbol{x}_j, T_1)$$

$$+ \int_{T_1}^{T_1+\Delta t} \left[ f_i(\boldsymbol{\theta_{UP}}(s)) + \left[ \left.\frac{\partial f_i}{\partial x_j}\right|_{\boldsymbol{x}=\boldsymbol{\theta_{UP}}(s)} \right] \boldsymbol{\gamma}(\Delta \boldsymbol{x}_j, s) + O\left(\sum \gamma_j(s)\gamma_k(s)\right) \right] ds$$

$$\underset{A7}{=} \theta_{UP,i}(T_1 + \Delta t) + \gamma_i(\Delta \boldsymbol{x}_j, T_1)$$

$$+ \int_{T_1}^{T_1+\Delta t} \left[ \left.\frac{\partial f_i}{\partial x_j}\right|_{\boldsymbol{x}=\boldsymbol{\theta_{UP}}(s)} \right] \boldsymbol{\gamma}(\Delta \boldsymbol{x}_j, s) ds + \int_{T_1}^{T_1+\Delta t} O\left(\sum \gamma_j(s)\gamma_k(s)\right) ds.$$

Then, this can be further rewritten as:

$$\theta_{P,i}(\Delta \boldsymbol{x}_j, T_1 + \Delta t) - \theta_{UP,i}(T_1 + \Delta t) - \gamma_i(\Delta \boldsymbol{x}_j, T_1)$$

$$= \int_{T_1}^{T_1+\Delta t} \left[ \left.\frac{\partial f_i}{\partial x_j}\right|_{\boldsymbol{x}=\boldsymbol{\theta_{UP}}(s)} \right] \boldsymbol{\gamma}(\Delta \boldsymbol{x}_j, s) ds + \int_{T_1}^{T_1+\Delta t} O\left(\sum \gamma_j(s)\gamma_k(s)\right) ds \qquad \text{(A8)}$$

From Eq. (A6), it gives:

$$\gamma_i(\Delta \boldsymbol{x}_j, T_1 + \Delta t) - \gamma_i(\Delta \boldsymbol{x}_j, T_1)$$

$$= \int_{T_1}^{T_1+\Delta t} \left[ \left.\frac{\partial f_i}{\partial x_j}\right|_{\boldsymbol{x}=\boldsymbol{\theta_{UP}}(s)} \right] \boldsymbol{\gamma}(\Delta \boldsymbol{x}_j, s) ds + \int_{T_1}^{T_1+\Delta t} O\left(\sum \gamma_j(s)\gamma_k(s)\right) ds \qquad (A9)$$

Dividing both sides of Eq. (A9) by $\Delta t$ and taking a limitation $\Delta t$ going to zero gives:

$$\left.\frac{d\gamma_i(\Delta \boldsymbol{x}_j, T)}{dT}\right|_{T=T_1} = \left[ \left.\frac{\partial f_i}{\partial x_j}\right|_{\boldsymbol{x}=\boldsymbol{\theta_{UP}}(T_1)} \right] \boldsymbol{\gamma}(\Delta \boldsymbol{x}_j, T_1) + O\left(\sum \gamma_j(T_1)\gamma_k(T_1)\right). \qquad (A10)$$

**APPENDIX 3.2** How to numerically solve the linearized ODE

To numerically solve the linearized ODE given by Eq. (20), we need to prepare the unperturbed orbit of the original nonlinear ODE model, $\varphi(\varphi(\boldsymbol{x}_0, k\tau), T), 0 < T < \tau$. Note that 4th-order Runge-Kutta method needs to evaluate the derivatives at $T + \frac{1}{2}\Delta t$ as well as $T$ and $T + \Delta t$ for obtaining the system state at $T + \Delta t$ from that at $T$, where $\Delta t$ is the time interval for numerical integrations. Therefore, we needed to prepare the numerical solution of the unperturbed orbit with the time interval $\frac{1}{2}\Delta t$ in $0 < T < \tau$, which results in greater computational costs than when directly evaluating the nonlinear ODEs with the time interval $\Delta t$.

**Appendix 4:** Open proposition for numerical stability

Regarding the numerical method in Eq. (21), one remark here is that the appropriate (maximum) value of $\Delta x$ depends on $\tau$. As shown in Eq. (18′) in **Appendix 2**, the parameter $\tau$ can be distinguished into $\tau_1$ as the data point interval and $\tau_2$ as the time scale for evaluating interaction strength. Then, Eqn (21) can be also generalized as:

$$S_{i,j,\tau_2}^{C,adj}(k\tau_1) \approx \frac{[\gamma_n(\tau_2, \Delta x)]_i}{\Delta x} - \delta_{ij}. \tag{A11}$$

When $\tau_2$ becomes large, $\Delta x$ should be smaller to have a good approximation of the limitation value. Although the condition in Eq. (A12) would hopefully be realized, the situation in Eq. (A13) could happen. This is related to the continuity of the solution of ODE in terms of initial conditions. Since we focus on the system that has chaotic attractors, it is not sure if the condition in Eq.(A12) is always satisfied. Only when Eq. (A12) is satisfied, the numerical approximation by Eq. (A11) can avoid numerical divergence to infinity, acting as the good approximation of the case $\Delta x \to 0$.

For any large $\tau_2$(> 0), there always exists enough small $\Delta x$ > 0, s.t.,

$$[\gamma_n(\tau_2, \Delta x)]_i = O(\Delta x), \tag{A12}$$

but the following is not satisfied

$$[\gamma_n(\tau_2, \Delta x)]_i = O(1). \tag{A13}$$

**APPENDIX 5** Partial derivatives for the ODE model

We have the following partial derivatives of the model in Eq. (26):

$$\frac{\partial f_{P_1}}{\partial P_1} = v_1\lambda_1 \frac{C_1}{C_1 + C_1^*} - v_1, \frac{\partial f_{P_1}}{\partial P_2} = 0, \frac{\partial f_{P_1}}{\partial C_1} = v_1\lambda_1 \frac{P_1 C_1^*}{(C_1 + C_1^*)^2}, \frac{\partial f_{P_1}}{\partial C_2} = \frac{\partial f_{P_1}}{\partial R} = 0,$$

$$\frac{\partial f_{P_2}}{\partial P_1} = 0, \frac{\partial f_{P_2}}{\partial P_2} = v_2\lambda_2 \frac{C_2}{C_2 + C_2^*} - v_1, \frac{\partial f_{P_2}}{\partial C_1} = 0, \frac{\partial f_{P_2}}{\partial C_2} = v_2\lambda_2 \frac{P_2 C_2^*}{(C_2 + C_2^*)^2}, \frac{\partial f_{P_2}}{\partial R} = 0,$$

$$\frac{\partial f_{C_1}}{\partial P_1} = -v_1\lambda_1 \frac{C_1}{C_1 + C_1^*}, \frac{\partial f_{C_1}}{\partial P_2} = 0, \frac{\partial f_{C_1}}{\partial C_1} = \mu_1\kappa_1 \frac{R}{R + R^*} - v_1\lambda_1 \frac{P_1 C_1^*}{(C_1 + C_1^*)^2} - \mu_1, \frac{\partial f_{C_1}}{\partial C_2} = 0,$$

$$\frac{\partial f_{C_1}}{\partial R} = \mu_1\kappa_1 \frac{C_1 R^*}{(R + R^*)^2},$$

$$\frac{\partial f_{C_2}}{\partial P_1} = 0, \frac{\partial f_{C_2}}{\partial P_2} = -v_2\lambda_2 \frac{C_1}{C_1 + C_1^*}, \frac{\partial f_{C_2}}{\partial C_1} = 0, \frac{\partial f_{C_2}}{\partial C_2} = \mu_2\kappa_2 \frac{R}{R + R^*} - v_2\lambda_2 \frac{P_2 C_2^*}{(C_2 + C_2^*)^2} - \mu_2,$$

$$\frac{\partial f_{C_2}}{\partial R} = \mu_2\kappa_2 \frac{C_2 R^*}{(R + R^*)^2},$$

$$\frac{\partial f_R}{\partial P_1} = \frac{\partial f_R}{\partial P_2} = 0, \frac{\partial f_R}{\partial C_1} = -\mu_1\kappa_1 \frac{R}{R + R^*}, \frac{\partial f_R}{\partial C_2} = -\mu_2\kappa_2 \frac{R}{R + R^*},$$

$$\frac{\partial f_R}{\partial R} = \left(1 - \frac{2}{k}\right) - \sum_{i=1,2} \mu_i\kappa_i \frac{C_i R^*}{(R + R^*)^2}.$$